\documentclass[usenatbib,twocolumn]{mnras}
\usepackage[utf8]{inputenc}

\usepackage{newtxtext,newtxmath}
\usepackage[T1]{fontenc}
\usepackage{ae,aecompl}
\usepackage[]{units}
\usepackage{graphicx}
\usepackage{todonotes}
\usepackage{amsmath}
\usepackage{threeparttable}
\usepackage{url}
\usepackage{hyperref}
\usepackage{bm}
\usepackage{color}
\usepackage{hhline}
\usepackage{subfigure}
\usepackage{lineno}
\usepackage{caption}

\title[Assembly bias]{Evidence for galaxy assembly bias in BOSS CMASS redshift-space galaxy correlation function}
\author[S. Yuan et al.]{
Sihan Yuan$^{1}$\thanks{E-mail: sihan.yuan@cfa.harvard.edu}, Boryana Hadzhiyska$^{1}$, Sownak Bose$^{1}$, Daniel J. Eisenstein$^{1}$, Hong Guo$^{2}$
\\
$^{1}$Center for Astrophysics | Harvard $\&$ Smithsonian, 60 Garden St., Cambridge, MA, 02138, USA \\
$^{2}$Key Laboratory for Research in Galaxies and Cosmology, Shanghai Astronomical Observatory, Shanghai 200030, China
}\date{December 2018}

\begin{document}

\maketitle

\label{firstpage}
\pagerange{\pageref{firstpage}--\pageref{lastpage}}

\begin{abstract} 
Building accurate and flexible galaxy-halo connection models is crucial in modeling galaxy clustering on non-linear scales. Recent studies have found that halo concentration by itself cannot capture the full galaxy assembly bias effect and that the local environment of the halo can be an excellent indicator of galaxy assembly bias. In this paper, we propose an extended halo occupation distribution model (HOD) that includes both a concentration-based assembly bias term and an environment-based assembly bias term. We use this model to achieve a good fit ($\chi^2/\mathrm{DoF} = 1.35$) on the 2D redshift-space 2-point correlation function (2PCF) of the Baryon Oscillation Spectroscopic Survey (BOSS) CMASS galaxy sample. We find that the inclusion of both assembly bias terms is strongly favored by the data and the standard 5-parameter HOD is strongly rejected. More interestingly, the redshift-space 2PCF drives the assembly bias parameters in a way that preferentially assigns galaxies to lower mass halos. This results in galaxy-galaxy lensing predictions that are within $1\sigma$ agreement with the observation, alleviating the perceived tension between galaxy clustering and lensing. We also showcase a consistent 3-5$\sigma$ preference for a positive environment-based assembly bias that persists over variations in the fit. We speculate that the environmental dependence might be driven by underlying processes such as mergers and feedback, but might also be indicative of a larger halo boundaries such as the splashback radius. Regardless, this work highlights the importance of building flexible galaxy-halo connection models and demonstrates the extra constraining power of the redshift-space 2PCF.
\end{abstract}
\begin{keywords}
cosmology: large-scale structure of Universe -- cosmology: dark matter -- galaxies: haloes -- gravitational lensing: weak -- methods: analytical  -- methods: statistical
\end{keywords}

\section{Introduction}

In the standard framework of structure formation in a $\Lambda$CDM universe, galaxies are predicted to form and evolve in dark matter halos \citep{1978White}. While the distribution and structure of dark matter halos is directly tied to the underlying cosmology, the extent to which galaxies do is less so. To extract cosmological information and understand galaxy formation from  observed galaxy clustering statistics, it is critical to correctly model the connection between galaxies and their underlying dark matter halos. The most popular model of the galaxy-halo connection is the Halo Occupation Distribution model \citep[HOD; e.g.][]{2000Peacock, 2001Scoccimarro, 2002Berlind, 2005Zheng, 2007bZheng}. The HOD makes the assumption that all galaxies live inside dark matter halos, and expresses the average number of galaxies contained within an an individual halo as a function of the halo mass. More specifically, the simplest formulation of the HOD assumes that galaxy occupation is determined {\it solely} by halo mass, an assumption that rests on the long-standing and widely accepted theoretical prediction that halo mass is the attribute that most strongly correlates with the halo abundance and halo clustering as well as the properties of the galaxies residing in it \citep[][]{1978White, 1984Blumenthal}. 

However, \citet{2005Gao, 2005Zentner, 2006Wechsler, 2007Gao, 2007Croton, 2008Li} showed that at fixed halo mass, halo clustering also depends on secondary halo properties that correlate with halo asssembly history, an effect known as \textit{halo assembly bias}.
Additionally, a series of studies employing hydrodynamical simulations and semi-analytic models have found clear evidence that galaxy occupation correlates with secondary halo properties beyond just halo mass \citep[e.g.][]{2006Zhu, 2018Artale, 2018Zehavi, 2019Bose, 2019Contreras, 2020Hadzhiyska, 2020Xu}. This phenomenon is commonly known as \textit{Galaxy Assembly Bias} (or assembly bias hereafter) to differentiate it from halo assembly bias. \citet{2018Wechsler} offers a more rigorous definition of assembly bias as: at fixed halo mass, the galaxy properties or number of galaxies within dark matter halos may depend on secondary halo properties that themselves show a halo assembly bias signature.
Ignoring the effects of assembly bias has been shown to introduce significant errors in inferring galaxy-halo connection models and bias galaxy formation models \citep{2014Zentner, 2014Pujol, 2019Lange}. 

Several studies have implemented ways to incorporate a secondary dependence into the HOD formalism \citep[e.g.][]{2015Paranjape, 2016Hearin, 2018McEwen, 2018Yuan, 2019bWibking, 2019Walsh, 2020Xu}. The different methodologies can be summarized into two approaches: one approach is to assign galaxies to halos according to a secondary property within each mass bin, such as the decorated HOD framework proposed in \citet{2016Hearin}. The other approach is to introduce a secondary dependence on the existing HOD parameter set, but where the definition of halo mass itself is modified by a dependence on the secondary parameter, such as in \citet{2018Yuan, 2019Walsh}. \citet{2019Zentner} applied the decorated HOD framework showed the first tentative observational evidence for galaxy assembly bias in SDSS DR7 main galaxy sample.

Thus far, halo concentration, which is a measure of how centrally-peaked the density profile of the halo is, has long been regarded as the standard secondary parameter to use in assembly bias studies. This choice is largely motivated by N-body simulations which show that halo concentration, along with several other parameters that correlate with the halo assembly history, are good predictors of halo assembly bias \citep{2006Wechsler, 2007Croton, 2018Mao}. However, recent studies that have systemically compared various secondary dependencies for assembly bias in hydrodynamical simulations and semi-analytic models have found that halo concentration only accounts for part of the actual assembly bias \citep{2020Hadzhiyska, 2020Xu}. Both studies find that the local environment of the halo at the present day is an excellent predictor of assembly bias, where the environment is defined as either the smoothed local matter density or the number of neighboring halos within some radius. We point out that the term ``assembly bias'' is somewhat of a misnomer in this context, as dependencies such as local environment identified today do not necessarily relate to the past assembly history of the halos. In this paper, we mean assembly bias in its most general sense, inclusive of all secondary dependencies in the galaxy-halo connection model. 

However, one must be careful when discussing halo environment in the context of assembly bias, as this may at first seem tautological. Since the environment of the halo is a statement on clustering itself, the halo environment cannot be naively used as the explanation for why there is assembly bias, since it is logically necessary that objects selected from dense environments exhibit stronger clustering. However, if one's objective is to reproduce realistic galaxy mocks rather than explaining it, then it is fair to use the halo environment as an indicator of assembly bias in decorating  galaxy-halo connection models. In fact, as \citet{2020Hadzhiyska} and \citet{2020Xu} have shown, halo environment is potentially the most effective indicator of assembly bias. In the context of dark matter only simulations, halo environment is also powerful because it is an easily accessible property, which bypasses the need to construct halo merger trees or resolve sub-halo structure. 

There is also growing evidence that  halo environment affects galaxy evolution and, therefore, the galaxy-halo connection directly. Both \citet{2020Hadzhiyska} and \citet{2020Xu} showed that while the environment is defined locally (typically between 1-5$h^{-1}$Mpc), it seems to capture the assembly bias effects on much larger scales, up to tens of megaparsecs. This suggests that the environment might trace underlying processes that, at least partially, drive assembly bias. This should not be surprising as excursion set theory predicts a correlation between the halo environment and its formation history \citep{1991Bond, 2007Zentner}. In the context of the cosmic web structure, studies have shown that, when controlled for halo mass, galaxy evolution depends on its proximity to close-by cosmic filaments, and the content and kinematics of neighboring gas reservoirs \citep[e.g.][]{2017Chen, 2017Poudel, 2018Laigle, 2019Salerno, 2019Kraljic, 2020Song}. \citet{2020Obuljen} found a 5$\sigma$ detection of an anisotropic assembly bias that correlates galaxy properties with large-scale tidal field in BOSS CMASS data. 
A series of papers including \citet{2017Tinker, 2018bTinker, 2018Tinker} studied the relation between various observed galaxy properties and environment at fixed halo mass and stellar mass and found that star-forming galaxies tend to live in underdense environments whereas quenched passive galaxies tend to occupy overdense environments. Independent studies such as \citet{2018Lee, 2018Dragomir, 2019Behroozi} found similar trends in simulations and in separate datasets. 

In this paper, we propose an extended HOD model that incorporates both a concentration-based assembly bias term and an environment-based assembly bias term. We constrain such an HOD with the observed two-dimensional galaxy redshift-space 2-point correlation function (2PCF) of the Baryon Oscillation Spectroscopic Survey \citep[BOSS,][]{2011Eisenstein, 2013Dawson} CMASS sample between $0.46\leq z \leq 0.6$ (Data Release 12). We show that the redshift-space 2PCF strongly prefers the inclusion of both assembly bias terms and affects the fit in a way that reduces the typical host halo mass of galaxies. We also show that the resulting best-fit HOD predicts the galaxy-galaxy lensing signal to within $1\sigma$, significantly reducing the perceived tension between galaxy clustering and lensing. Our results show that incorporating various galaxy assembly bias effects is an important ingredient in an accurate and flexible HOD model, and that the perceived tension between galaxy clustering and g-g lensing might partially be due to over-simplistic HOD models and the lack of constraining power of the projected 2PCF. We also showcase a consistent preference for a positive environment-based assembly bias by the data. This serves as further evidence that an environment-based assembly bias, together with the traditional concentration-based assembly bias, should be included in future galaxy-halo connection models. 

The paper is organized as follows: In section~\ref{sec:theory}, we describe the extended HOD framework and the implementation of assembly bias parameters. In section~\ref{sec:data}, we describe the observed redshift-space 2PCF and the simulation sets we employ in this work. In section~\ref{sec:method}, we present the HOD fitting methodology and the optimizations developed for our analysis. In section~\ref{sec:results}, we showcase our HOD fits with and without the assembly bias terms, the corresponding lensing predictions, and the best-fit value of the environment-based assembly bias across variations to the fit.  In section~\ref{sec:discussion}, we discuss the parameter recovery of our routine, alternative environment definitions, and compare our assembly bias fit to previous works. We conclude in section~\ref{sec:conclusion}.

\section{The extended HOD framework}
\label{sec:theory}





The HOD is a popular empirical framework used to populate dark matter halos with central and satellite galaxies as a function of halo mass. However, given the simplistic nature of galaxy assignment, this model may also be a source of systematic errors for cosmological applications. In this section, we briefly review the baseline HOD formalism and discuss physically motivated extensions to the standard HOD. A more detailed description of the HOD and some of the extensions discussed here can be found in \citet{2018Yuan} and in section 2 of \citet{2019Yuan}.
Our extended HOD code is publicly available as the \textsc{GRAND-HOD} package\footnote{\url{https://github.com/SandyYuan/GRAND-HOD}}.

\subsection{The baseline model}

The baseline 5-parameter HOD model \citep{2007Zheng} predicts the mean number of central galaxies and satellite galaxies as a function of halo mass $M$:
\begin{align}
&\bar{n}_{\mathrm{cent}}(M) = \frac{1}{2}\mathrm{erfc} \left[\frac{\ln(M_{\mathrm{cut}}/M)}{\sqrt{2}\sigma}\right], \nonumber \\
& \bar{n}_{\textrm{sat}}(M) = \left[\frac{M-\kappa M_{\textrm{cut}}}{M_1}\right]^{\alpha}\bar{n}_{\mathrm{cent}}(M),
\label{equ:standard_hod}
\end{align}
where halo mass is defined as the mass contained within a radius encompassing 200
times the background density, $M_{200\textrm{b}}$.
The five parameters of this model are $M_{\textrm{cut}}, M_1, \sigma, \alpha, \kappa$. $M_{\textrm{cut}}$ characterizes the minimum halo mass to host a central galaxy. $M_1$ characterizes the typical halo mass that hosts one satellite galaxy. $\sigma$ describes the steepness of the transition from 0 to 1 in the number of central galaxies. $\alpha$ is the power law index on the number of satellite galaxies. $\kappa M_\textrm{cut}$ gives the minimum halo mass to host a satellite galaxy. The actual number of central galaxies follows a Bernoulli distribution while the actual number of satellite galaxies follows a Poisson distribution. The spatial distribution of satellites within the halo follows the halo matter density distribution. Traditionally, the satellites in each halo are distributed according to a Navarro–Frenk–White profile (NFW) profile \citep{1997Navarro}. Instead, we distribute the satellite positions according to the particle sub-sample of each halo. While this method incurs a computational cost, this approach has the benefit of preserving the shape and dynamics of the dark-matter halo. For the baseline HOD, we simply give each particle of the halo equal probability of hosting a satellite galaxy, and the satellite galaxy inherits the velocity of the host particle. 

Finally, to generate galaxies positions in redshift-space, we assume the $z$ axis to be the Line-Of-Sight (LOS) and modify the $z$ coordinates of the galaxies according to: 
\begin{equation}
    X'_z = X_z + \frac{v_z}{H}(1+z),
\end{equation}
where $X'_z$ is the redshift-space comoving position of the galaxy, and $X_z$ is the real-space comoving position of the galaxy. $v_z$ is the $z$-component of the galaxy velocity.

In the following sub-sections, we introduce 6 additional physically-motivated HOD parameters.

\subsection{Satellite profile parameters $s$ and $s_p$}

We first introduce two parameters that allow for flexibility in the radial distribution of satellite galaxies within each halo. 

The radial distance parameter, $s$, deviates the satellite spatial distribution away from the halo matter density profile by giving preference to particles based on their radial distance to halo center. A positive $s$ preferentially situates satellites on the outskirts of the halo, whereas a negative $s$ preferentially situates satellites towards the inner region of the halo. Figure~2 of \citet{2018Yuan} shows how changing $s$ affects the mock 2PCF. The range of $s$ is defined to be between $-1$ and 1. The $s$ parameter is motivated by baryonic processes that can bias the concentration of baryons within the dark matter potential well \citep[e.g.][]{2010Duffy, 2010Abadi, 2017Chua, 2017Peirani}.

The perihelion distance parameter, $s_p$, is related to the $s$ parameter but additionally folds in the velocity information of the particles. Specifically, it gives preference to particles based on their perihelion distance to the halo center, i.e. their closest-approach distance to the halo center given their current trajectory. The perihelion distance is calculated by solving the Kepler equations within an NFW potential well \citep[Equations~6--9 in][]{2018Yuan}. The impact of $s_p$ on the 2PCF is shown in Figure~5 of \citet{2018Yuan}. This parameter is motivated by processes such as ram-pressure stripping and tidal disruption. 

\subsection{Velocity bias parameters $s_v$ and $\alpha_c$}

Another important set of parameters we employ in order to accurately model the redshift-space correlation function are the satellite and central velocity bias parameters \citep[e.g.][]{2003Berlind, 2003Yoshikawa, 2005vdBosch, 2011Skibba, 2015aGuo}.

First, we define the satellite velocity bias parameter, $s_v$, which biases the satellite velocity distribution away from that of the host halo. A positive $s_v$ preferentially assigns satellites to high peculiar velocity particles of the halo, and vice versa. We note that our implementation lets the satellites assume the peculiar velocity of the their underlying matter field, thus guaranteeing that the satellite galaxies still obey Newtonian physics in the halo potential. This is in contrast to existing velocity bias implementations where satellite velocities are increased/decreased without altering their positions, breaking Newtonian physics. 
A key difference between these two approaches is that our velocity bias implementation has a small effect on the projected correlation function, whereas existing implementations have strictly zero impact on the projection clustering. Figure~4 of \citet{2018Yuan} shows how $s_v$ affects the predicted 2PCF. The range of $s_v$ is defined to be between $-1$ and 1. 

We also introduce the central velocity bias parameter, $\alpha_c$. In the baseline implementation, the central galaxy is assumed to have the position and velocity of the halo center-of-mass (CoM). When invoking velocity bias, the central galaxy velocity is given by:
\begin{equation}
    v_\mathrm{cent, z} = v_\mathrm{CoM, z} + \delta v(\alpha_c \sigma_{\mathrm{LOS}}),
    \label{equ:alphac}
\end{equation}
where $v_\mathrm{cent, z}$ is the LOS velocity of the central, and $v_\mathrm{CoM, z}$ is the LOS velocity of the halo CoM. $\sigma_{\mathrm{LOS}}$ is the LOS velocity dispersion of the halo particles. $\alpha_c$ is the central velocity bias parameter. $\delta v$ is drawn from a Gaussian distribution with zero mean and standard deviation of $\alpha_c \sigma_{\mathrm{LOS}}$. The central velocity bias has strictly no effect on projected clustering, but affects the ``length'' of the finger-of-god in redshift-space. While $\alpha_c$ can technically vary between 0 and $+\infty$, we expect the true $\alpha_c$ to be no greater than 1. 

\subsection{Assembly bias parameters $A$ and $A_e$}

So far, all our extensions to the baseline HOD have only dealt with the distribution of galaxies within each halo while respecting the assumption that the number of galaxies depends only on halo mass. In this section, we relax this assumption by introducing two secondary dependencies: halo concentration and halo environment. 

We first define the concentration-based assembly bias parameter, $A$. The motivation for using halo concentration as the secondary parameter is that it is correlated with the formation histories of dark matter halos, with earlier forming halos having higher concentrations at fixed halo mass \citep{2002Wechsler, 2003Zhao, 2006Wechsler, 2009Zhao, 2017Villarreal}.
In this work, we define halo concentration as:
\begin{equation}
    c = \frac{r_{\textrm{vir}}}{r_{s, \textrm{Klypin}}},
    \label{equ:concentration}
\end{equation}
where $r_{\textrm{vir}}$ is the virial radius of the halo and $r_{s, \textrm{Klypin}}$ is the velocity-based Klypin scale radius \citep{2011Klypin}.
Our assembly bias implementation is based on a routine that preserves the overall galaxy number density, first described in \citet{2018Yuan}. To summarize briefly, we rank all halos by their mass, and compute the corresponding list of expected number of galaxies using the baseline HOD that depends only on mass. Then, we perturb the ranking of halos by defining a pseudo-mass: 
\begin{equation}
    \log_{10} M_\mathrm{pseudo}  = \log_{10} M + A\delta_c,
    \label{equ:A}
\end{equation}
where $\delta_c = (c - \bar{c}(M))/\sigma_c(M)$ is the halo concentration subtracted by the mean concentration in that specific mass bin and normalized by the corresponding scatter in concentration. 
We perturb the halo ranking by sorting by $M_\mathrm{pseudo}$ and then map the unperturbed list of expected number of galaxies onto the perturbed list of halos bijectively. Our implementation essentially swaps the galaxies between halos while preserving the total number density of galaxies.

However, it does not preserve the expected number of galaxies for a given halo mass $\langle \bar{n}_g|M\rangle$, in contrast to the assembly bias implementation in the \textsc{Halotools} decorated HOD framework \citep{2013Behroozi, 2016Hearin}. A detailed description of our implementation can be found in section~3 of \citet{2018Yuan}. 
Figure~6 of \citet{2018Yuan} shows the effect of $A$ on the predicted 2PCF. The range of $A$ is technically between $-\infty$ and $\infty$, but we expect $A$ to be between -1 and 1. 

Similarly, we define the environmental assembly bias parameter, $A_e$, which incorporates the halo environment as the secondary dependence. To define halo environment, we adopt the same formalism as \citet{2020Hadzhiyska}. Specifically, for each halo, we find all neighboring halos (including subhalos) beyond its virial radius but within $r_\mathrm{max} = 5 h^{-1}$Mpc of the halo center. We sum the mass of all these neighboring halos as $M_{\mathrm{env}}$, and we compute the environment factor, $f_{\mathrm{env}}$, as:
\begin{equation}
    f_\mathrm{env} = M_\mathrm{env} / \bar{M}_\mathrm{env}(M),
    \label{equ:fenv}
\end{equation}
where $\bar{M}_\mathrm{env}(M)$ is the mean environment factor within halo mass bin $M$. Finally, we incorporate $f_\mathrm{env}$ into the pseudo-mass definition in Equation~\ref{equ:A} and introduce the environmental assembly bias parameter $A_e$:
\begin{equation}
    \log_{10} M_\mathrm{pseudo}  = \log_{10} M + A\delta_c + A_e f_\mathrm{env}.
    \label{equ:Ae}
\end{equation}
Again, we incorporate these assembly bias effects into our HOD by re-ranking the halos with $M_\mathrm{pseudo}$ to essentially swap galaxies between halos of different concentration and environment. The choice of $r_\mathrm{max}$ is meant to be large enough to capture the immediate vicinity of the halo without extending deep into the 2-halo regime. We revisit this definition in Section~\ref{sec:discussion}.

It is important to note that while our implementation has the distinct advantage of incorporating multiple secondary dependencies, our assembly bias model is also limited by the fact that we do not distinguish between assembly biases for the central galaxies and satellite galaxies, as exemplified in some earlier assembly bias frameworks \citep[e.g.][]{2016Hearin, 2020Xu}. Recent simulation-based works have also found evidence that centrals and satellites might indeed have distinct assembly bias signatures \citep[e.g.][]{2019Bose}. For this work, we do not make this distinction for model simplicity and to limit the number of necessary parameters. We explore alternative assembly bias models that distinguish between centrals and satellites in upcoming work.

\subsection{Incompleteness factor $f_\mathrm{ic}$}
The final parameter in our extended HOD model is the incompleteness factor \citep[e.g.][]{2016Torres, 2016Leauthaud, 2018Guo}. The inclusion of incompleteness is partially motivated by detection of incompleteness in the BOSS CMASS and LOWZ galaxy samples compared with theoretical stellar mass functions, but it is also needed to marginalize over uncertainties in the galaxy number densities since we do not model its full redshift dependence. We define $f_\mathrm{ic}$ as a modification to $\bar{n}_{\mathrm{cent}}$ in Equation~\ref{equ:standard_hod}:
\begin{align}
&\bar{n}_{\mathrm{cent}} = \frac{f_\mathrm{ic}}{2}\mathrm{erfc} \left[\frac{\ln(M_{\mathrm{cut}}/M)}{\sqrt{2}\sigma}\right],
\label{equ:fic}
\end{align}
where $0 < f_\mathrm{ic} \leq 1$. Our implementation simply uniformly downsamples the mock galaxies by $f_\mathrm{ic}$ to produce the desired galaxy number density. \hfill \break

To summarize, our extended HOD model contains the 5 baseline parameters $M_{\textrm{cut}}, M_1, \sigma, \alpha, \kappa$, 6 extended parameters $s, s_p, s_v, \alpha_c, A, Ae$, and 1 incompleteness factor $f_\mathrm{ic}$, for a total of 12 parameters.

\section{Data and Simulations}
\label{sec:data}

\subsection{BOSS CMASS galaxy sample}
\label{sbsec:cmass}
The Baryon Oscillation Spectroscopic Survey \citep[BOSS; ][]{2012Bolton, 2013Dawson} is part of the SDSS-III programme \citep{2011Eisenstein}. BOSS Data Release 12 (DR12) provides redshifts for 1.5 million galaxies in an effective area of 9329 square degrees divided into two samples: LOWZ and CMASS. The LOWZ galaxies are selected to be the brightest and reddest of the low-redshift galaxy population at $z < 0.4$, whereas the CMASS sample is designed to isolate galaxies of approximately constant mass at higher redshift ($z > 0.4$), most of them being also Luminous Red Galaxies \citep[LRGs,][]{2016Reid, 2016Torres}. The survey footprint is divided into chunks which are covered in overlapping plates of radius $\sim 1.49$ degrees. Each plate can house up to 1000 fibres, but due to the finite size of the fibre housing, no two fibres can be placed closer than $62$ arcsec, referred to as the fibre collision scale \citep{2012Guo}. 

\begin{figure}
    \centering
    \hspace*{-0.6cm}
    \includegraphics[width = 3.5in]{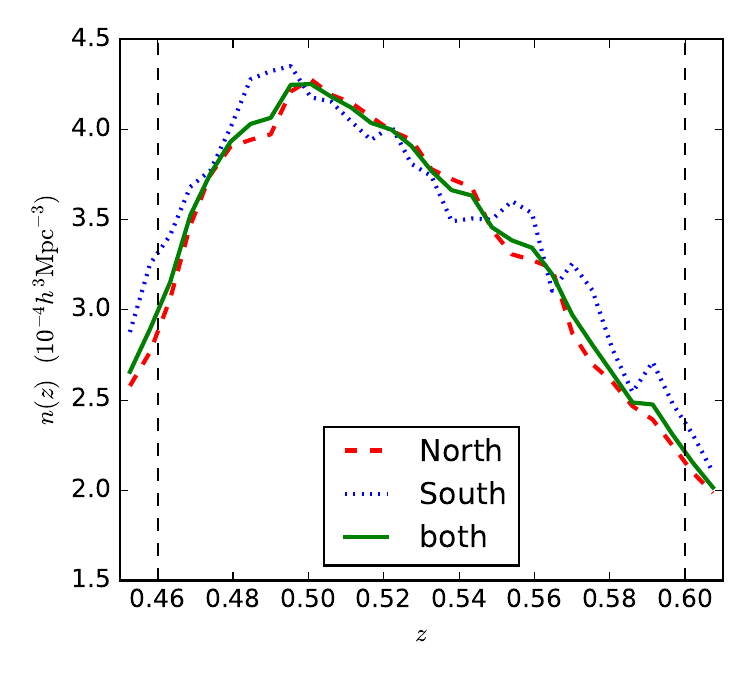}
    \vspace{-0.3cm}
    \caption{The CMASS DR12 galaxy comoving number density distribution across our redshift range. The red dashed line corresponds to the north galactic cap, whereas the blue dotted line corresponds to the south galactic cap. The green solid line shows the combined number density. The two vertical lines mark $z = 0.46$ and $z = 0.6$, respectively. }
    \label{fig:nz}
\end{figure}

For this paper, we limit our measurements to the galaxy sample between redshift $0.46 < z < 0.6$ in DR12. We choose this moderate redshift range for completeness and to minimize the systematics due to redshift evolution. Applying this redshift range to both the north and south galactic caps gives a total of approximately 600,000 galaxies in our sample. We showcase the number density variation over redshift in Figure~\ref{fig:nz}. The average galaxy number density is given by $n_\mathrm{data} = (3.01\pm 0.03)\times 10^{-4} h^{3}$Mpc$^{-3}$. 

\begin{figure*}
    \centering
    \hspace*{-0.6cm}
    \includegraphics[width = 6.5in]{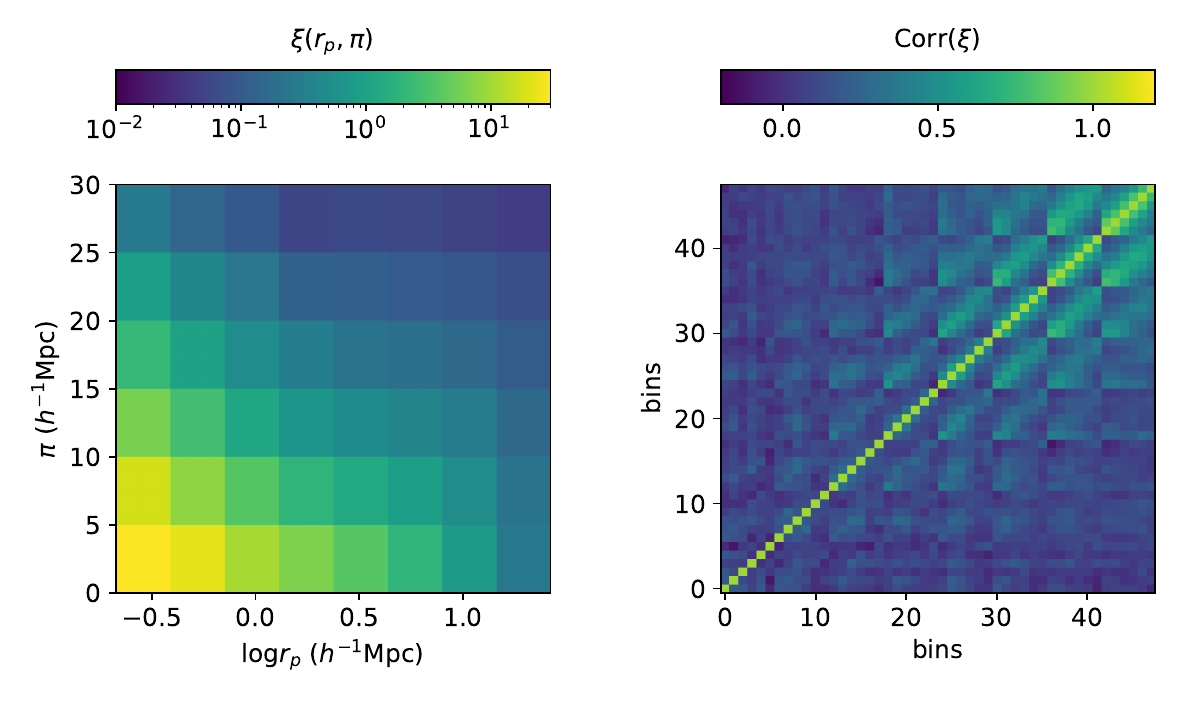}
    \vspace{-0.3cm}
    \caption{The redshift-space 2-point correlation function (left) of the BOSS CMASS DR12 galaxies at $0.46 < z < 0.6$ and its corresponding covariance matrix (right). $r_p$ is the transverse comoving distance between galaxies. $\pi$ is the LOS comoving distance between galaxies. The right hand side shows the correlation matrix,  calculated from 400 jackknife sub-samples. The bins are formed by flattening the $\xi$ bins column-by-column, with large bin number corresponding to large $r_p$. For example, the bin 0-5 correspond to the first $r_p$ bin but increasing $\pi$.}
    \label{fig:xi_boss}
\end{figure*}

Figure~\ref{fig:xi_boss} shows the redshift-space 2PCF of the same BOSS sample and its corresponding correlation matrix, assuming Planck 2015 cosmology. 
The redshift-space 2PCF $\xi(r_p, \pi)$ is computed using the \citet{1993Landy} estimator:
\begin{equation}
    \xi(r_p, \pi) = \frac{DD - 2DR + RR}{RR},
    \label{equ:xi_def}
\end{equation}
where $DD$, $DR$, and $RR$ are the normalized numbers of data-data, data-random, and random-random pair counts in each bin of $(r_p, \pi)$, and $r_p$ and $\pi$ are transverse and line-of-sight (LOS) separations in comoving units. For this paper, we choose a coarse binning to ensure reasonable accuracy on the covariance matrix, with 8 logarithmically-spaced bins between 0.169$h^{-1}$Mpc and 30$h^{-1}$Mpc in the transverse direction, and 6 linearly-spaced bins between 0 and 30$h^{-1}$Mpc bins along the LOS direction. 

We have corrected the fibre collision effect following the method of \cite{2012Guo}, by separating galaxies into collided and decollided populations and assuming those collided galaxies with measured redshifts in the plate-overlap regions are representative of the overall collided population. The final corrected correlation function can be obtained by summing up the contributions from the two populations.

The correlation matrix is defined relative to the covariance matrix as $\mathrm{Corr}(\xi)_{ij} = \mathrm{Cov}(\xi)_{ij}/\sqrt{\mathrm{Cov}(\xi)_{ii}\mathrm{Cov}(\xi)_{jj}}$. The covariance matrix is calculated from 400 jackknife samples. The $x$ and $y$ axes of the right panel in Figure~\ref{fig:xi_boss} show the same bins as on the left panel, flattened in a column-by-column fashion such that the transverse separation $r_p$ increases with bin number. Overall, we see that the off-diagonal power is relatively small at small transverse scales and becomes more significant at larger transverse scales. This suggests that the error at small $r_p$ is shot-noise dominated, while sample variance becomes dominant at large $r_p$. 

\subsection{N-body simulations and halo finders}

For the primary results of this paper, we generate our galaxy mocks from the \textsc{AbacusCosmos} N-body simulation suite, generated by the fast and high-precision \textsc{Abacus} N-body code \citep[][Ferrer et al., in preparation; Metchnik $\&$ Pinto, in preparation]{2018Garrison, 2016Garrison}\footnote{For more details, see \url{https://lgarrison.github.io/AbacusCosmos}.}. We use 20 boxes of comoving size 1100~$h^{-1}$Mpc with Planck 2015 cosmology \citep{2016Planck} at redshift $z = 0.5$. We quote the cosmology parameter values as $\Omega_c h^2 = 0.1199$, $\Omega_b h^2 = 0.02222$, $\sigma_8 = 0.830$, $n_s = 0.9652$, $h = 0.6726$, and $w_0 = -1$. These boxes are set to different initial phases to generate unique outputs. Each box contains 1440$^3$ dark matter particles of mass $4\times 10^{10}$ $h^{-1}M_{\odot}$. The force softening length is 0.06~$h^{-1}$~Mpc. Dark matter halos are found and characterized using the {\sc Rockstar} \citep{2013Behroozi} halo finder. 

When testing the environmental assembly bias in Section~\ref{subsec:Ae}, we also incorporate the \textsc{AbacusSummit} simulation suite, which is a set of large, high-accuracy cosmological N-body simulations designed to meet the Cosmological Simulation Requirements of the Dark Energy Spectroscopic Instrument (DESI) survey \citep{2013arXiv1308.0847L}. \textsc{AbacusSummit} consists of over 140 simulations most of which contain $6912^3$ particles within a 2$h^{-1}$Gpc volume, which yields a particle mass of $2.1 \times 10^9 h^{-1}M_\odot$. \footnote{For more details, see \url{https://abacussummit.readthedocs.io/en/latest/abacussummit.html}}

The \textsc{AbacusSummit} boxes that we employ include one with the primary Planck 2018 $\Lambda$CDM cosmology as the benchmark and 4 other boxes with perturbed cosmologies, specifically varying cosmological parameters $\Omega_c$ and $\sigma_8$. The Planck 2018 cosmology parameters are $\Omega_c h^2 = 0.1200$, $\Omega_b h^2 = 0.02237$, $\sigma_8 = 0.811355$, $n_s = 0.9649$, $h = 0.6736$, $w_0 = -1$, and $w_a = 0$. Note that this is a slightly different cosmology than the Planck 2016 cosmology used for the \textsc{AbacusCosmos} simulations, the largest difference being a $\sim 2\%$ smaller $\sigma_8$ in the new cosmology.
We concentrate our study of these simulations at redshift $z = 0.5$ and make use of the output data products including particle subsamples, {\sc CompaSO} (Hadzhiyska et al. 2020 in prep.) and {\sc Rockstar} halo catalogs.

The two simulation suites also utilize different halo finders. The \textsc{AbacusCosmos} simulations use {\sc Rockstar} whereas the \textsc{AbacusSummit} simulations use {\sc CompaSO}.

{\sc Rockstar} is a temporal, phase-space 
halo finder considered to be highly accurate in
determining particle-halo membership, as it 
uses information about both the phase space distribution of the
particles as well as their temporal evolution  \citep{2013Behroozi}. 
This is because
having information about the relative motion
of two halos makes the process of finding tidal remnants
and determining halo boundaries
substantially more effective, while having temporal information
helps to maximize the consistency of halo properties 
across time, rather than just within a single snapshot.

The {\sc CompaSO} halo-finding algorithm was designed
for the \textsc{AbacusSummit} suite of 
high-performance cosmological N-body simulations,
as a highly efficient on-the-fly group 
finder (Hadzhiyska et al. 2020, in prep.). 
{\sc CompaSO} builds on the existing 
spherical overdensity (SO) algorithm
by taking into consideration the tidal radius
around a smaller halo before competitively
assigning halo membership to the particles
in an effort to more effectively deblend halos.
Among other features, the {\sc CompaSO} finder also
allows for the formation of new halos on the 
outskirts of growing halos, which alleviates
a known issue of configuration-space halo 
finders of failing to identify halos close to
the centers of larger halos. 

A detailed comparison between the two halo finders will be made in Hadzhiyska et al. in prep.

\section{Methods}
\label{sec:method}

The fundamental technological challenge is to search a high-dimensional extended HOD parameter space for a prescription that best reproduces the observed redshift-space 2PCF signal and galaxy number density. Perhaps the most popular approach for fitting an extended HOD model on data is to employ a so-called emulator. The emulator approach is where we train a surrogate model of the observable within a HOD training set before we fit the best-fit surrogate model on data. An example of an HOD emulator with the extended HOD model in this work was implemented in \citet{2019Yuan}. However, we found that the emulator for such a high dimensional parameter space has poor accuracy and breaks down outside the training range. The quadratic emulator model we employed was also not flexible enough for a wide training range to enable a comprehensive search. Thus, in this paper, we opt for a direct global optimization of the likelihood function to find the optimal HOD. A direct optimization has a much wider range in parameter space and is not limited by the shape of the surrogate model in an emulator. However, it requires repeatedly generating mock galaxy catalogs from simulations, which can be prohibitively expensive, especially since we are utilizing the particle catalog in addition to the halo catalog in this analysis.

In this section, we describe our likelihood function, its maximization, and then the key methods implemented to accelerate the HOD code.

\subsection{The maximum-likelihood routine}
\label{sec:method_hod}

We assume Gaussian likelihood and express the log-likelhood using the chi-square technique. The $\chi^2$ is given in two parts, corresponding to errors on the redshift-space 2PCF and errors on the galaxy number density:
\begin{equation}
\chi^2  = \chi^2_{\xi} + \chi^2_{n_g},
\label{equ:logL}
\end{equation}
where
\begin{equation}
       \chi^2_{\xi}  = (\bm{\xi}_{\mathrm{mock}} - \bm{\xi}_{\mathrm{data}})^T \bm{C}^{-1}(\bm{\xi}_{\mathrm{mock}} - \bm{\xi}_{\mathrm{data}}),
       \label{equ:chi2xi}
\end{equation}
and 
\begin{equation}
   \chi^2_{n_g} = \begin{cases}
   \left(\frac{n_{\mathrm{mock}} - n_{\mathrm{data}}}{\sigma_{n}/5}\right)^2 & (n_{\mathrm{mock}} < n_{\mathrm{data}}) \\
   \left(\frac{n_{\mathrm{data}}(1/f_{\mathrm{ic}} - 1)}{\sigma_{n}}\right)^2 & (n_{\mathrm{mock}} \geq n_{\mathrm{data}}).
   \end{cases}
   \label{equ:chi2ng}
\end{equation}
We define $\bm{C}$ as the jackknife covariance matrix on $\xi$, and $\sigma_n$ is the jackknife uncertainty of the galaxy number density. The $\chi^2_{n_g}$ is an asymmetric normal around the observed number density $n_\mathrm{data}$. When the mock number density is higher than data number density $(n_{\mathrm{mock}} \geq n_{\mathrm{data}})$, we invoke the incompleteness fraction $f_{\mathrm{ic}}$ that uniformly downsamples the mock galaxies to match the data number density. In this case, we penalize $f_\mathrm{ic}$ values that are far from 1. When the mock number density is less than the data number density $(n_{\mathrm{mock}} < n_{\mathrm{data}})$, we give a much steeper penalty on the difference between $(n_{\mathrm{mock}}$ and $n_{\mathrm{data}})$. This definition of $\chi^2_{n_g}$ allows for modest incompleteness in the observed galaxy sample while penalizing HOD models that produce insufficient galaxy number density or too many galaxies. 
For the rest of this paper, we set $n_\mathrm{data} = 3.0\times 10^{-4} h^{3}$Mpc$^{-3}$ and $\sigma_n = 4.0\times 10^{-5} h^{3}$Mpc$^{-3}$. Note that we choose a more lenient $\sigma_n$ than the jackknife uncertainty on galaxy number density in section~\ref{sbsec:cmass} because we want to explore a larger HOD parameter space. 

To minimize $\chi^2$, we utilize a global optimization algorithm known as covariance matrix adaptation evolution strategy \citep[CMA-ES,][]{2001Hansen}. CMA-ES is an evolutionary algorithm that stochastically varies and selects on a school of candidate solutions, resembling the evolution of a biological system. In the simplest terms, the algorithm works by updating the mean and covariance matrix of the distribution to increase the probability of previously successful search vectors in each step until the candidate solutions converge to the global optimum. The algorithm also records and exploits the time-wise history of the search for faster stepping while also preventing premature convergence. The specific implementation of CMA-ES we use is part of the publicly available StochOPy (STOCHastic OPtimization for PYthon) package. \footnote{\url{https://github.com/keurfonluu/StochOPy}.} 
To assess the error bars on the best fit, we run 22 Markov chain Monte Carlo (MCMC) chains initialized around the best-fit with the \textsc{emcee} package \citep{2013Foreman}. We quote our $1\sigma$ error bars as the standard deviation of the marginalized posterior distribution. 

\subsection{Accelerating the HOD code}

The key challenge in our HOD global optimization is speed. Each HOD evaluation requires generating mock galaxies on the halo catalogs and then computing summary statistics. The first step is particularly time-consuming since we are using a particle based HOD. In the following paragraphs, we describe several key speed-ups to enable a fast particle-based HOD implementation. 

The first and most obvious speed-up comes from parallelization. Given that generating mock galaxies is a so-called ``embarrassingly parallel'' problem on the halo level, the performance gain scales roughly linearly with the number of CPU cores utilized, given a sufficient amount of memory and I/O bandwidth. All computation in this analysis is done on a custom-built desktop, where we distribute the computation over 20 cores on a pair of Intel Xeon E5-2630v4 CPU clocked at 2.2 Ghz for a roughly $20\times$ performance gain.

Another ${\sim}20\times$ speed-up comes from utilizing a \textsc{Numba} just-in-time (jit) compiler \citep{cite_numba}, which converts slow \textsc{Python} code to fast machine code. \textsc{Numba} is especially powerful for long loops of un-vectorized code, which is the case for generating mock satellite galaxies. Note that generating satellites is the main performance bottleneck as generating central galaxies does not query the particle subsamples. The \textsc{Numba} compiler brings the time to generate the satellites down to ${\sim10}\times$ that of the centrals.

The third speed-up comes from pre-downsampling the halos and particles for satellite generation. The idea here is that satellites are rare compared to centrals, especially at halo mass $<10^{14}$. Thus, we aggressively downsample the halos at smaller halo mass and correspondingly upweight the expected number of satellites in each of the downsampled halos. This way, we significantly reduce the number of halos looped over in each HOD evaluation without losing fidelity, yielding a significant performance improvement. Similarly, we can downsample the particles in the halos to further increase performance. With our final choice of downsampling functions, we achieve another $3\times$ speed-up in our HOD evaluation. We suspect that our downsampling is still relatively conservative, and an even greater speed-up is attainable. 

I/O is another performance bottleneck. We largely overcome this by pre-loading the downsampled halo and particle file on memory before we start an optimization chain. To control memory usage, we only load relevant halo and particle information. Since the extended HOD parameters ($s$, $s_p$, $s_v$) only interact with the ranking of particle properties within each halo instead of the particle properties themselves, we can pre-compute these ranks and load them on memory. Another potential slowdown is when the HOD generates too many galaxies. To avoid this problem, we calculate the incompleteness factor $f_\mathrm{ic}$ to match the observed galaxy number density before we start an HOD evaluation. Then we scale down the number of centrals and satellites for each halo prior to assigning galaxies. 

Finally, we compute the redshift-space 2PCF using the high-performance \textsc{Corrfunc} code in parallel \citep{2020Sinha}. This specific computation turns out to be fast compared to generating mock galaxies. For the \textsc{AbacusCosmos} simulations, our final optimized pipeline, given our machine specifications, evaluates a new HOD over a ${\sim}10 h^{-1}$Gpc$^3$ volume and computes its $\chi^2$ in roughly 7 seconds, 5 seconds of which are spent on generating mocks and 2 seconds on computing the 2PCF. The performance is similar for the \textsc{AbacusSummit} simulations. For the rest of this paper, we fit the HOD with only the first 8 of the 20 \textsc{AbacusCosmos} simulation boxes, due to limitations in system memory. For the \textsc{AbacusSummit} fits, we only use one $2h^{-1}$Gpc box at each cosmology, which is equivalent to approximately 6 \textsc{AbacusCosmos} boxes in volume. 

\section{Results}
\label{sec:results}

\begin{table*}
\centering
\begin{tabular}{ c | c c c c c }
\hhline {======}
Parameter name & Prior [min, max] & $\xi$ fit with $A$ and $A_e$ & $\xi$ fit with $A$ & $\xi$ fit with $A_e$ & $\xi$ fit with neither \\ 
\hline
$\log_{10}(M_{\textrm{cut}}/h^{-1}M_\odot)$ & [12.5, 14] & $13.33\pm 0.03$ & $13.35\pm 0.03$ & $13.37\pm 0.04$ & $13.16\pm 0.02$ \\ 
\\[-1em]
$\log_{10}(M_1/h^{-1}M_\odot)$ & / & $14.47\pm 0.03$ & $14.52\pm 0.02$ & $14.33\pm 0.03$ & $14.34\pm 0.02$\\
\\[-1em]
$\sigma$ & [0.1, 2.0] & $0.61\pm 0.05$ & $0.52 \pm 0.07$ & $0.94\pm 0.06$ & $0.11\pm 0.09$\\
\\[-1em]
$\alpha$ & [0.7, 1.5] & $1.32\pm 0.05$ & $1.39 \pm 0.04$ & $1.01\pm 0.04$ & $1.16\pm 0.04$\\
\\[-1em]
$\kappa$ & [0.1, 2.0] & $0.2\pm 0.1$ & $0.1\pm 0.1$ & $0.2\pm 0.1$ & $0.2\pm 0.1$\\
\hline
$s$ & [-1.0, 1.0] & $0.1\pm 0.1$ & $0.0\pm 0.1$ & $0.3\pm 0.1$ & $0.6\pm  0.1$\\
\\[-1em]
$s_v$ & [-1.0, 1.0] & $0.8\pm 0.1$ & $0.6\pm 0.1$ & $0.1\pm 0.1$ & $0.5\pm 0.1$\\
\\[-1em]
$\alpha_c$ & [0.0, 2.0] &  $0.22\pm 0.03$ & $0.26\pm 0.01$ & $0.07\pm 0.07$ & $0.21\pm 0.02$\\
\\[-1em]
$s_p$ & [-1.0, 1.0] & $-1.0\pm 0.1$ & $-0.9\pm 0.2$ & $-1.0\pm 0.1$ & $-1.0\pm 0.1$\\
\\[-1em]
$A$ & [-1.0, 1.0] &  $-0.88\pm 0.08$ & $-0.79\pm 0.06$ & / & /\\
\\[-1em]
$A_e$ & [-1.0, 1.0] & $0.040\pm 0.009$ & / & $0.032\pm 0.05$ & / \\
\hline 
$f_{\mathrm{ic}}$ & / & 0.91 & 1.00 & 0.74 & 0.67 \\\hline 
Final $\chi^2$ (DoF) & / & 50 (37) & 67 (38) & 73 (38) & 93 (39)\\
BIC & / & 92 & 107  & 111  & 128 \\
\hline 
$\log_{10}\bar{M}_h/M_\odot$ & / & 13.52 & 13.56 & 13.57 & 13.61\\
\hline 
\end{tabular} 
\caption{Summary of the key HOD fits in this study. The first column lists the HOD parameters, incompleteness factor $f_{\mathrm{ic}}$, the final $\chi^2$, degree-of-freedom (DoF), and the average halo mass per galaxy $\log_{10}\bar{M}_h/M_\odot$. The second column shows the prior constraints. The next 4 columns summarize the 4 key fits of this study. The prior constraints on $\log M_1$ are not listed because we choose to constrain the satellite fraction $0 < f_\mathrm{sat} < 0.2$ instead. Note there is a one-to-one correspondence between $M_1$ and $f_\mathrm{sat}$ when the other HOD parameters are fixed. The corresponding best-fit $f_\mathrm{sat} = 9.4\%$. The errors shown are $1\sigma$ marginalized errors.}
\label{tab:bestfits}
\end{table*}

Table~\ref{tab:bestfits} summarizes the four key fits of this study. All four fits are constrained on the observed $\xi(r_p, \pi)$ and the observed galaxy number density. 
The four fits are identical except for which assembly terms are included. The first fit includes both $A$ and $A_e$. The second and third fits only use one assembly bias term, $A$ and $A_e$, respectively. The fourth fit includes neither. The comparison of these four fits lets us assess the importance of each assembly bias term. When a fit does not include an assembly bias parameter, we simply fix that parameter to 0. Besides the HOD parameters, we also list the best-fit $\chi^2$, the degree-of-freedom (DoF), the Bayesian information criterion (BIC), and the average halo mass per galaxy. The $\chi^2$ shown has been corrected for the finite simulation volume, and also corrected for the covariance matrix inversion bias following \citet{2007Hartlap}. The limits of the tophat prior constraints on the parameters are listed in the second column. The prior constraints on $\log M_1$ are not shown because we choose to constrain the satellite fraction $0 < f_\mathrm{sat} < 0.2$ instead.

We define the average halo mass per galaxy simply as
\begin{equation}
    \bar{M}_h = \frac{\sum_g M_h}{N_g},
    \label{equ:avgM}
\end{equation}
where the numerator sums the halo masses of all galaxies in the mock, and $N_g$ is the total number of galaxies in the mock. 
The average halo mass per galaxy characterizes the typical halo mass of a galaxy given an HOD prescription, which is interesting in assessing the effect of assembly bias and indicative of the galaxy-galaxy lensing prediction. 

\subsection{$\xi(r_p, \pi)$ fit with both $A$ and $A_e$}

The first (leftmost) HOD fit listed in Table~\ref{tab:bestfits} uses the full set of extended HOD parameters including both assembly bias terms. We achieve a good fit on $\xi(r_p, \pi)$, with a final $\chi^2 = 50$ (DoF = 37) and a BIC of 92. For comparison, if we fit the $\xi(r_p, \pi)$ with just the standard 5-parameter HOD with no extensions, then we get $\chi^2 = 151$ and a BIC of 174. This shows that the standard 5-parameter HOD is strongly disfavored by the redshift-space correlation function. 

\begin{figure*}
    \centering
    \hspace*{-1cm}
    \includegraphics[width = 8in]{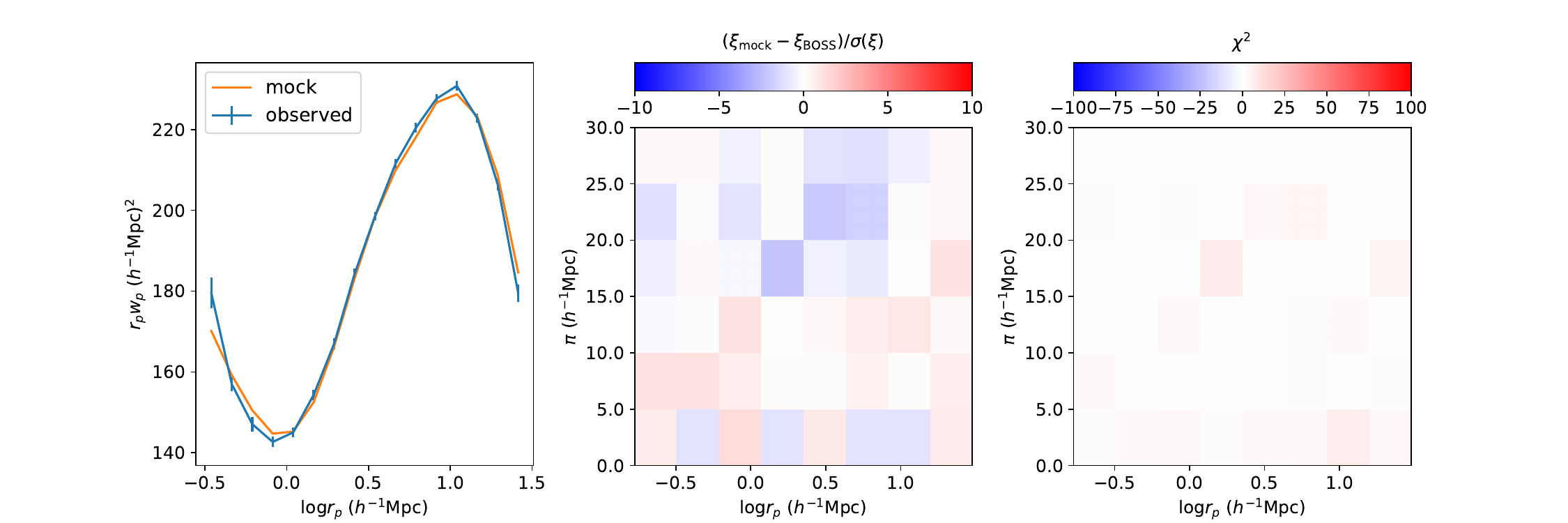}
    \hspace*{-1cm}
    \caption{The best-fit 2PCF using \textsc{AbacusCosmos} boxes, showing the $\xi(r_p, \pi)$ fit with both assembly bias terms $A$ and $A_e$. The left panel compares the projected 2PCF of the best-fit HOD (orange) to that of the data (blue). The error bars on the data are computed from the diagonal of the inverted covariance matrix. The middle panel compares the best-fit $\xi(r_p, \pi)$ with the data, where the errors, $\sigma(\xi)$, are computed from the diagonal of the inverse covariance matrix. The right panel showcases the contribution to the final $\chi^2$ from each bin. }
    \label{fig:xi_bestfit}
\end{figure*}

Figure~\ref{fig:xi_bestfit} visualizes the corresponding best-fit 2PCF. The left panel shows the best-fit projected 2PCF in orange and the observation in blue. The middle panel shows the normalized difference between the best-fit $\xi(r_p, \pi)$ and the observation. The normalization $\sigma(\xi)$ is derived from of the diagonal of the inverse covariance matrix, i.e. $\sigma = 1/\sqrt{\mathrm{diag}(\bm{C^{-1}})}$. The right panel shows the $\chi^2$ contribution from each bin, computed by multiplying array $(\bm{\xi}_{\mathrm{mock}} - \bm{\xi}_{\mathrm{data}})$ with array $\bm{C}^{-1}(\bm{\xi}_{\mathrm{mock}} - \bm{\xi}_{\mathrm{data}})$ term-wise. Summing over these terms gives the final $\chi^2_\xi$, as in Equation~\ref{equ:chi2xi}. 

We see that the best-fit HOD does provide a good fit to the projected 2PCF and the redshfit-space 2PCF. While a few bins in $\xi(r_p, \pi)$ seem to exhibit higher normalized error in the middle panel, it is important to note that the error bars $\sigma(\xi)$ are computed from the diagonal of the inverse covariane matrix, underestimating the true uncertainty due to the high off-diagonal power in the covariance matrix (refer to the right panel Figure~\ref{fig:xi_boss}). Thus, the actual difference between the mock and the data in these bins is less significant than it appears. The right panel incorporates the full covariance matrix and is thus more informative in judging the consistency between data and mock in each bin. One bin (column 4 row 4) stands out as it contributes 7 to the final $\chi^2$. Excluding this bin from the HOD fit does not meaningfully change the final parameter values.

To take a closer look at the best-fit parameter values, let us first take a look at the assembly bias parameters.
The concentration-based assembly bias parameter $A$ varies somewhat depending on the details of the fit, but generally yields rather negative best-fit values from -0.5 to -0.9. The variation is likely affected by degeneracies within the HOD space and multimodality in the likelihood surface, in which case the errorbar shown is underestimated. The negative $A$ has the effect of moving galaxies into less massive and puffier halos. In terms of its clustering signature, a negative $A$ increases the projected clustering at intermediate and large transverse scales ($r_p > 1 h^{-1}$Mpc) while reducing clustering at large LOS separations, especially at small transverse scales ($r_p < 1 h^{-1}$Mpc), suggesting a smaller velocity dispersion. \citet{2019Lange} also shows degeneracies between $A$ and cosmology, specifically $f\sigma_8$. Thus, a negative $A$ could also indicate that our presumed $f\sigma_8$ is too high. Unfortunately, we do not have a sufficiently large range in the cosmologies probed by our simulations to properly verify this possibility. Thus, we leave it as an opportunity for a future study. 

The environment-based assembly bias parameter $A_e$ yields a best-fit value of $0.04$ and is stable across different fits, suggesting that galaxies preferentially populate halos in denser environments. We point out that the relative magnitudes of $A$ and $A_e$ parameters are deceptive as the actual amplitude of the effect depends on that value of the normalized concentration $\delta_c$ and the environment factor $f_\mathrm{env}$ (Equation~\ref{equ:A} and \ref{equ:Ae}). The actual contribution of $A_e = 0.04$ and $A = -0.8$ on $\xi(r_p, \pi)$ are comparable in amplitude. 

The other extended parameters also yield interesting fits. 
$s$, which sets the radial distribution of satellite galaxies, is slightly positive but consistent with zero. We find a large variation in the best-fit $s_v$, anywhere from 0.2 to 0.8, depending on the details of the fit. This variation is again likely due to degeneracies between $s_v$ and some other levers in the extended HOD, leading to multimodality in the likelihood surface. Regardless, the positive $s_v$ has the effect of increasing satellite velocities and thus increasing the finger-of-god effet. Unlike $s_v$, the best-fit for the central velocity bias parameter $\alpha_c$ is stable at ${\sim}0.2$ across different fits. This is consistent with the multipole fits in \citet{2015aGuo}. The best-fit value of $s_p$ is also consistently close to $-1$ across our fits, suggesting that the observation strongly favors to put some satellites on highly eccentric orbits that pass through the central regions of the halo. $s_p$ is a novel addition to the HOD, and its significance possibly indicates the existence of a subset of infalling or splashback galaxies in the CMASS sample. This is an interesting result and we reserve a more detailed discussion on $s_p$ in a separate paper. 


\subsection{$\xi(r_p, \pi)$ fit without both $A$ and $A_e$}

The second and third fit in Table~\ref{tab:bestfits} only incorporate one assembly bias term, $A$ and $A_e$, respectively. The fourth fit includes neither assembly bias terms. 
Compared to the first fit with both $A$ and $A_e$, the second and third fit yields significantly worse $\chi^2$, with large increase to the BIC, $\Delta$BIC = 15 and $\Delta$BIC = 19, respectively. Conversely, comparing to the fit with no assembly bias, the inclusion of either $A$ or $A_e$ leads to significantly better fits. Comparing the first fit and the fourth fit, we see that the inclusion of both assembly bias term is strongly favored by the data, with a $\Delta$BIC = -36. Thus, we conclude that the redshift-space 2PCF calls for the simultaneous inclusion of both a concentration-based assembly bias and an environmental assembly bias, at least in our HOD framework at Planck 2015 cosmology.

\begin{figure}
    \centering
    \hspace*{-0.5cm}
    \includegraphics[width = 2.5in]{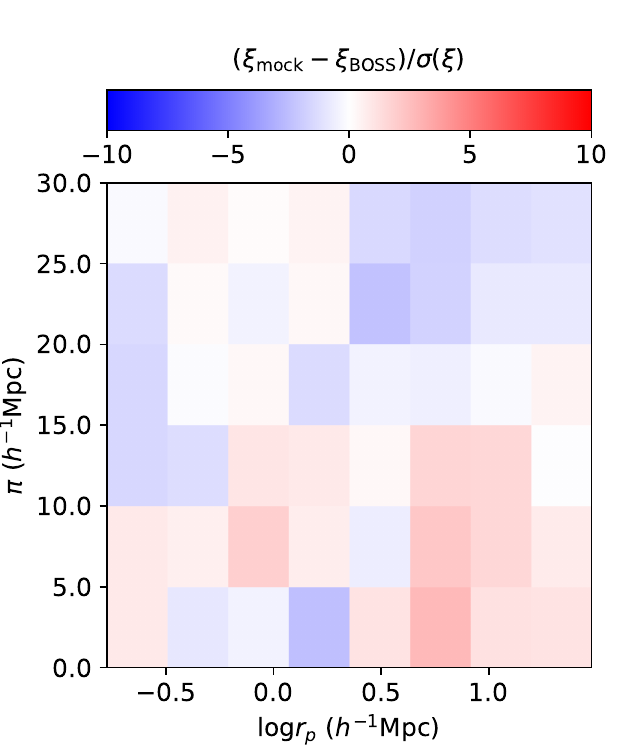}
    \caption{The residual in the redshift-space 2PCF $\xi(r_p, \pi)$ relative to the data when neither assembly biases are included in the HOD, corresponding to the last best fit listed in Table~\ref{tab:bestfits}. Comparing to the fit including both assembly biases (middle panel of Figure~\ref{fig:xi_bestfit}), we see significantly larger residuals at large $r_p$, and worse prediction of the LOS structure.}
    \label{fig:xi_compare}
\end{figure}

Figure~\ref{fig:xi_compare} shows the residuals in the redshift-space 2PCF $\xi(r_p, \pi)$ compared to the data when neither assembly biases are included in the fit, i.e. the last fit shown in Table~\ref{tab:bestfits}. Compared to the middle panel of Figure~\ref{fig:xi_bestfit}, where we show the residuals of the HOD fit including both assembly biases, we see that the residuals here are noticeably larger, especially at large $r_p\sim$5-10$h^{-1}$Mpc. The LOS structure reproduced by the no assembly bias fit is also worse, across all $r_p$. To explain this, note that on the one hand, the concentration-based assembly bias $A$ has a strong effect on the galaxy velocity dispersion as it moves galaxies into more or less massive halos. Thus, $A$ is strongly sensitive to the LOS structure of $\xi(r_p, \pi)$, at small and large $r_p$. This explains why the inclusion of $A$ improves the fit on the LOS structure of $\xi(r_p, \pi)$. The environmental assembly bias $A_e$, on the other hand, does not produce as strong a LOS signature, but it predominantly affects the projected clustering on intermediate scales $r_p \sim$2-10$h^{-1}$Mpc. This explains why the inclusion of $A_e$ in the model reduces the residuals at $r_p\sim$5-10$h^{-1}$Mpc. This comparison between Figure~\ref{fig:xi_compare} and the middle panel of Figure~\ref{fig:xi_bestfit} directly showcases how the inclusion of the two assembly bias terms improve the $\xi(r_p, \pi)$ fit and highlights their importance in a flexible HOD model.

In terms of the average halo mass per galaxy, we see that the inclusion of either assembly bias results in a 10-12$\%$ decrease compared to the fit with no assembly bias, whereas the inclusion of both terms results in a $23\%$ decrease. In comparison, a 5-parameter HOD plus parameters $s$, $A$, and $A_e$ constrained on the projected correlation function $w_p$ and galaxy number density gives an average mass of $\bar{M}_h = 10^{13.62}M_\odot$, $26\%$ larger than that of the $\xi(r_p,\pi)$ fit with both assembly bias turned on. This indicates that the redshift-space 2PCF prefers to assign galaxies to lower mass halos at fixed bias, and the two assembly bias terms give the HOD flexibility to do so, allowing for a much better fit. The projected 2PCF does not exhibit this preference, even when modeled with both assembly biases turned on. This result highlights the extra constraining power offered by the LOS structure of the redshift-space 2PCF. The decrease in typical halo mass for galaxies also has significant implications for the galaxy-galaxy lensing signal, as we will discuss in the following sub-section. 

While we should not over-interpret the final parameter values of these fits due to the poor $\chi^2$, we can still compare the parameter values across these fits to gain intuition on what exactly is driving the fit. When $A$ is included, the fit prefers a strongly negative value, which has the effect of increasing large scale clustering while decreasing the typical halo mass of galaxies. However, a negative $A$ also results in a smaller velocity dispersion on the very small scale due to the less massive and puffier halos. To compensate for this effect, the fit chooses a larger $\log_{10} M_1$, $\alpha$, and $s_v$, which increases the finger-of-god effect on the small scale. In other words, it seems that the strong clustering amplitude at large scales is driving galaxies into less massive halos with a negative $A$, and then the satellite distribution parameters are then tuned to match the small-scale finger-of-god signature.  

The inclusion of $A_e$ has a similar effect, where a positive $A_e$ increases clustering on larger scales while decreasing the typical halo mass of galaxies. However, compared to $A$, the clustering signature of $A_e$ is more dependent on $r_p$, specifically its signature is strongest in the $1h^{-1}\mathrm{Mpc} < r_p < 4h^{-1}\mathrm{Mpc}$ range and weakens beyond that, whereas the signature of $A$ remains strong up to $30h^{-1}\mathrm{Mpc}$. $A_e$ also produces a weaker gradient along the LOS and has a rather small signature on the small scale ($r_p < 1h^{-1}\mathrm{Mpc}$) compared to $A$. Thus, the inclusion of $A_e$ triggers less of response from the parameters that control the satellites and the velocity dispersion, but it further decreases the average halo mass while boosting intermediate to large scale clustering. 

\subsection{The galaxy-galaxy lensing prediction}

\begin{figure*}
    \centering
    \hspace*{-0.6cm}
    \includegraphics[width = 7in]{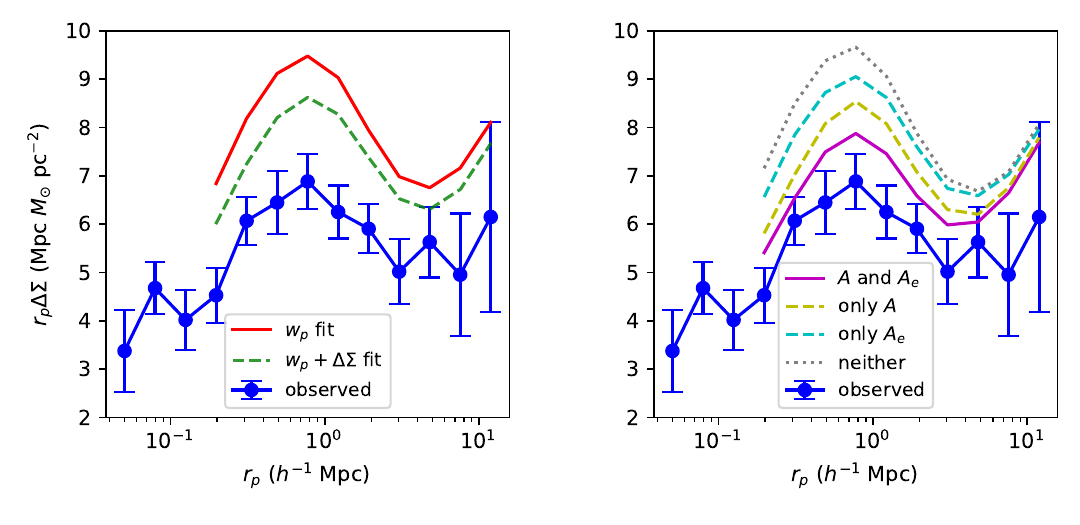}
    \vspace{-0.3cm}
    \caption{The comparison between the predicted galaxy-galaxy lensing signal and the observed galaxy-galaxy lensing signal. In both panels, the blue curve shows the observed lensing signal of the CMASS galaxy sample, quoted directly from \citet{2017Leauthaud}. In the left panel, the solid red curve shows the predicted lensing signal if we constrain the HOD with just the projected correlation function $w_p$ and the galaxy number density. The dashed green line shows the $w_p + \Delta \Sigma$ emulator fit from \citet{2019Yuan}. In the right panel, the magenta curve shows the predicted lensing signal of our standard $\xi(r_p, \pi)$ fit, including both $A$ and $A_e$. The dashed yellow (cyan) curve shows the prediction when we fit the $\xi(r_p, \pi)$ with only $A$ ($A_e$). The gray dotted line shows the prediction when we do not inlude either assembly biases in the HOD model. We see that by fitting the full redshift-space correlation function and incorporating both assembly biases into the HOD, we significantly reduce the tension between data and predictions, down to about $1\sigma$ level.}
    \label{fig:lensing}
\end{figure*}

A well known tension exists between galaxy clustering and galaxy-galaxy lensing (g-g lensing). \citet{2017Leauthaud} found discrepancies of 20-40$\%$ between their measurements of g-g lensing for CMASS galaxies and a model predicted from mock galaxy catalogs generated at Planck cosmology that match the CMASS projected correlation function \citep[][see Figure~7 of \citealt{2017Leauthaud}]{2014Reid, 2016Saito}. \citet{2019Lange} extended this result by finding a similar ${\sim} 25\%$ discrepancy between the projected clustering measurement and the g-g lensing measurement in the BOSS LOWZ sample. In \citet{2019Yuan}, we reaffirmed this tension by fitting simultaneously the projected galaxy clustering and g-g lensing with an extended the HOD incorporating a concentration-based assembly bias prescription. The left panel of Figure~\ref{fig:lensing} reproduces this tension, where the blue curve showcases the observed lensing signal of the CMASS sample, and the red curve shows the prediction of a HOD constrained on the projected 2PCF. The dashed green curve indicates the joint fit from \citet{2019Yuan}. 
The best-fit HOD corresponding to the red curve achieves a very good fit of the projected 2PCF ($\chi^2 = 5.6$ and DoF = 9), with $<1\%$ error across all bins. However, it provides a very poor lensing prediction, approximately 20-40$\%$ larger than the observation. The joint fit of the projected 2PCF $w_p$ and the galaxy-galaxy lensing shown in the dashed green curve fails to fit either observable well, resulting in a 10-20$\%$ discrepancy with the observed $w_p$ (shown in Figure~4 of \citet{2019Yuan}) while reducing the lensing discrepancy by ${\sim}$10$\%$, not enough to reconcile with the observation. 

The inclusion of both assembly bias terms ($A$ and $A_e$) and the switch to redshift-space 2PCF presents an opportunity at reducing this tension. As we have shown, both assembly bias terms allow the redshift-space 2PCF to drive the fit in the direction of reducing the typical halo mass of galaxies. The right panel of Figure~\ref{fig:lensing} shows the g-g lensing predictions of our redshift-space clustering fits. Again, the solid blue curve shows the measurement from \citet{2017Leauthaud} on the CMASS sample. The solid magenta line represents the $\xi(r_p, \pi)$ fit with both $A$ and $A_e$, whereas the the dashed yellow and cyan curves show the fit with just $A$ and $A_e$, respectively. The dotted grey curve shows the fit with neither assembly bias terms. As expected, the introduction of both assembly bias terms leads to lower predicted lensing signal by assigning galaxies to less massive halos. 

The prediction of the $\xi(r_p, \pi)$ fit with both $A$ and $A_e$ matches the observation to within 1$\sigma$. This is a significant improvement compared to the $3\sigma$ discrepancies found with previous model predictions that fit the projected 2PCF $w_p$ using more simplistic HOD models \citep[e.g.][]{2014Reid, 2016Saito, 2016Torres, 2017bAlam, 2019Lange, 2019Yuan}. Comparing the predictions on the right panel, we see that the decrease in the lensing signal really comes from a combination of incorporating assembly biases and constraining on the redshift-space 2PCF. Without the assembly bias terms, the $\xi(r_p, \pi)$ fit actually shows no improvement over the $w_p$ fit. The flexibility introduced by the assembly bias terms is what allows for a good $\xi(r_p, \pi)$ fit, which drives down the typical halo mass of galaxies, thus decreasing the lensing signal. Comparing the ``with $A$'' fit on the right with the ``$w_p + \Delta\Sigma$ fit'' on the left, we see that the inclusion of just a concentration-based assembly bias results in a similar lensing prediction, even though one is constrained on $\xi(r_p, \pi)$ and the other is constrained on $w_p+\Delta\Sigma$. However, it is clear that the just a concentration-based assembly bias term is not sufficient in predicting the observed lensing signal. The combination of $A$ and $A_e$ is what uniquely brings the lensing prediction in agreement with the observation while also providing a good fit to $\xi(r_p, \pi)$.

It is interesting that the inclusion of $A$ brings the g-g lensing prediction into better consistency with data than the inclusion of $A_e$. This hierarchy between $A$ and $A_e$ echoes with the fact that the inclusion of $A$ yields a somewhat lower BIC and a slightly better fit to $\xi(r_p, \pi)$ than the inclusion of $A_e$ in Table~\ref{tab:bestfits}. This could mean that while both secondary dependencies are required to yield consistent predictions with data, the secondary dependency on concentration is more important than the dependency on environment in producing a more realistic HOD.

While we have shown that the combination of $A$ and $A_e$ is powerful in modeling both redshift-space clustering and g-g lensing, we do not claim that we have found the silver bullet to resolving the lensing tension. However, we believe that this represents a promising path towards reducing the lensing tension. A full solution of the lensing tension will likely also appeal to better handling of systematics and possibly small corrections in cosmology. We also do not claim that we have found the true prescription of galaxy assembly bias or the correct HOD model. Nevertheless, our findings highlight the importance of constructing flexible galaxy-halo connection models and employing more informative clustering statistics such as the redshift-space 2PCF. This result echoes the findings of \citet{2020Zu}, where the author found a sophisticated HOD model with detailed treatments of selection effects could reconcile the g-g lensing tension. However, it is argued that their model triggers an unrealistic satellite fraction \citep{2020Lange}. 
\citet{2020Amodeo} recently constrained cluster gas dynamics using Sunyaev-Zeldovich effect and found an excess non-thermal pressure due to baryonic processes that would reduce the lensing tension by $50\%$. These energetic ejection processes towards larger halo radii are consistent with the positive environmental dependency that we found, highlighting the importance of baryonic structure beyond the typical virial radius of the halo.

\subsection{Investigating the environmental assembly bias $A_e$}
\label{subsec:Ae}

\begin{figure}
    \centering
    \hspace*{-0.6cm}
    \includegraphics[width = 3.7in]{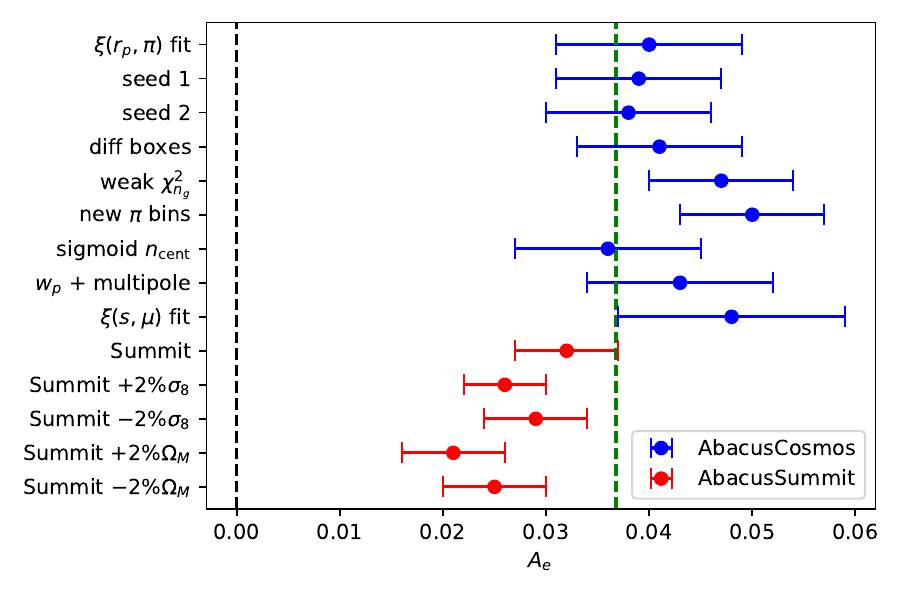}
    \vspace{-0.3cm}
    \caption{The best-fit values of the environmental assembly bias parameter $A_e$ across various fits. The blue markers represent fits using the \textsc{AbacusCosmos} simulations, whereas the red markers represent fits using the \textsc{AbacusSummit} simulations. The error bars are marginalized $1\sigma$ error bars. The dashed green line represents the average best-fit value across all the fits.}
    \label{fig:Aefits}
\end{figure}

\begin{figure}
    \centering
    \hspace*{-0.6cm}
    \includegraphics[width = 3.5in]{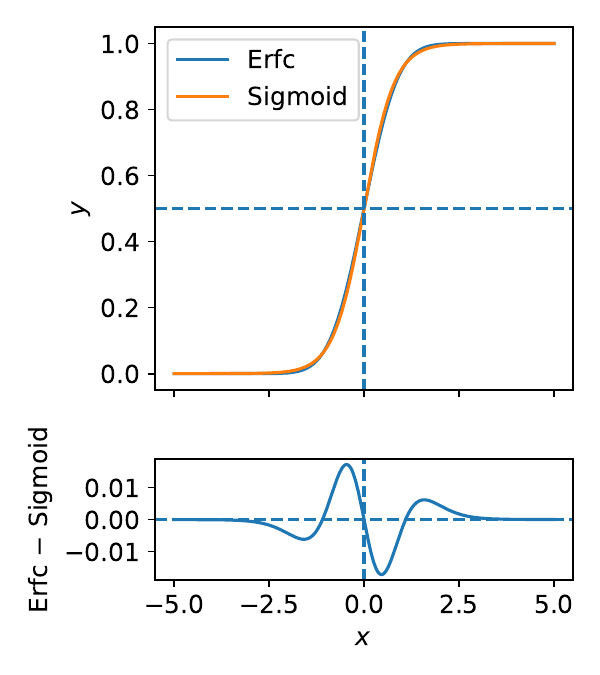}
    \vspace{-0.3cm}
    \caption{The top panel shows an example of a sigmoid function compared to a similar error function. The specific functions shown here are $0.5\mathrm{erfc}(x)$ in blue and $\mathrm{sigmoid}(2.5x)$ in orange. The bottom panel shows the difference between the two functions. Overall, the sigmoid produces a steeper incline in the middle but has a slower convergence to 1. }
    \label{fig:sigmoid}
\end{figure}

The novel environmental assembly bias parameter deserves some special attention as we showed that it behaves rather differently than the concentration-based assembly bias parameter and that it is indispensable in modeling the redshift-space 2PCF and predicting the observed lensing signal. 
We find further support for its legitimacy in the fact that its best-fit value is remarkably stable across all our fits, despite variations to the HOD, likelihood function, compression of the data, and modest perturbations to the assumed cosmology. We visualize its best-fit values across all these variations in Figure~\ref{fig:Aefits}, where the blue markers represent fits using the \textsc{AbacusCosmos} simulations, and the red markers represent fits using the \textsc{AbacusSummit} simulations. We briefly describe each of these fits as follows:
\begin{itemize}
    \item $\xi(r_p, \pi)$ fit: The $\xi(r_p, \pi)$ fit with both $A$ and $A_e$ as shown in Table~\ref{tab:bestfits}, using 8 \textsc{AbacusCosmos} boxes at Planck 2015 cosmology. 
    \item seed 1-2: Same as $\xi(r_p, \pi)$ fit, except using different random number seeds to marginalize over shot noise effects. 
    \item diff boxes:  Same as $\xi(r_p, \pi)$ fit, except using a different set of 8 of the 20 simulation boxes to marginalize over sample variance effects.
    \item weak $\chi^2_{n_g}$: Weakening the $n_g$ component of the likelihood function. Specifically, we modify $\chi^2_{n_g}$ as defined by Equation~\ref{equ:chi2ng} to essentially a step function:
    \begin{equation}
    \chi^2_{n_g} = \begin{cases}
    \left(\frac{n_{\mathrm{mock}} - n_{\mathrm{data}}}{\sigma_{n}}\right)^2 & (n_{\mathrm{mock}} < n_{\mathrm{data}}) \\
     0 & (n_{\mathrm{mock}} \geq n_{\mathrm{data}}),
     \end{cases}
    \label{equ:chi2ng_step}
    \end{equation}
    where $\sigma_n \approx 3\times 10^{-6} h^{3}$Mpc$^{-3}$ is the jackknife uncertainty on the observed $n_g$. This new $\chi^2_{n_g}$ does not penalize the HOD for producing too many galaxies, allowing for rather low incompleteness factors. The best-fit yields an incompleteness factor of $f_\mathrm{ic} = 0.67$.
    \item new $\pi$ bins: Nonlinear binning along the LOS direction to give more weight to the very small scales ${\sim}1h^{-1}$Mpc. The new bins are $\pi = [0, 0.5, 1, 5, 10, 20, 30]$. 
    \item sigmoid $n_{\mathrm{cent}}$: Using a sigmoid function instead of an error function for $\bar{n}_\mathrm{cent}$. This test is done to address concerns that the error function was chosen arbitrarily for the HOD and might not correctly represent the physical central galaxy occupation distribution. The sigmoid function serves as an alternative ``switch'' function from 0 to 1, prividing a ``softer'' ramp-up relative to the error function. Figure~\ref{fig:sigmoid} showcases the difference between a pair of similar sigmoid and error functions. We find that the best-fit $A_e$ is consistent with that of the error function fits. We also find no significant preference for using either the error function or the sigmoid function in the HOD.
    \item $w_p$ + multipole: Fitting $w_p + \xi_0 + \xi_2$ instead of $\xi(r_p, \pi)$. This fit and the next fit test whether different compressions of the redshift-space 2PCF affects the best-fit $A_e$. 
    \item $\xi(s, \mu)$ fit: Fitting $\xi(s, \mu)$ instead of $\xi(r_p, \pi)$. $s$ and $\mu$ essentially represent a polar coordinate system in the pair separation space, where $s$ is the scalar separation between the two galaxies, and $\mu$ denotes the angle between the pair separation vector and the LOS. 
    \item Summit: $\xi(r_p, \pi)$ fit using one \textsc{AbacusSummit} box at Planck 2018 cosmology. The halos are identified using the {\sc CompaSO} halo finder instead of {\sc Rockstar}.
    \item Summit $+2\%\sigma_8$: Same as Summit fit, but with $2\%$ larger $\sigma_8$ in the assumed cosmology. 
    \item Summit $-2\%\sigma_8$: Same as Summit fit, but with $2\%$ lower $\sigma_8$ in the assumed cosmology. 
    \item Summit $+2\%\Omega_M$: Same as Summit fit, but with $2\%$ larger $\Omega_M$ in the assumed cosmology. 
    \item Summit $-2\%\Omega_M$: Same as Summit fit, but with $2\%$ lower $\Omega_M$ in the assumed cosmology. 
\end{itemize}

We see that within the \textsc{AbacusCosmos} fits in blue, all the $A_e$ values are consistent within $1\sigma$, regardless of all the variations we introduced. The $A_e$ values for the \textsc{AbacusSummit} fits in red are more dispersed, likely caused by changes in cosmology and larger uncertainty due to the smaller simulation volume at perturbed cosmologies. The shift in $A_e$ between the two sets of simulations are likely due to a combination of the slightly different cosmology (Planck 2015 vs 2018), different N-body codes, and different halo finders ({\sc Rockstar} vs {\sc CompaSO}). Regardless, the consistent 3-5$\sigma$ preference for a positive $A_e$ indicates that the environment dependence is a relatively standalone effect independent of concentration-based assembly bias and other HOD parameters, at least in our HOD framework. The positive $A_e$ value with reasonable signal-to-noise is also consistent with the environmental assembly bias signatures found in hydrodynamical simulations \citep{2020Hadzhiyska, 2020bHadzhiyska, 2020Xu}. Finally, we note that while we find that the environmental dependence is important, we do not yet understand why it is so. We make a few attempts at gaining intuition on this phenomenon in section~\ref{subsec:rmax} and section~\ref{subsec:other_studies}. 

\section{Discussion}
\label{sec:discussion}
\subsection{Testing HOD parameter recovery}
To verify the effectiveness of our fitting procedure, we apply the fitting procedure to a mock redshift-space 2PCF generated from the simulations using a fiducial extended HOD. We also generate the jackknife covariance matrix from the same set of mocks, normalized to the BOSS CMASS volume. In the first test, we generate the mock observed 2PCF from the same set of 8 simulation boxes that are then used to do the fitting. This avoids the effects of cosmic variance, and only tests whether our optimization routine is capable of recovering the correct underlying HOD. We find excellent recovery of all HOD parameters, with errors typically $<1\%$. The parameter $\kappa$ has the worst recovery, with errors of a few percent.

In the second test, we generate the mock observed 2PCF from 8 different simulation boxes than the ones used for fitting, thus introducing sample variance. We repeat this test for different fiducial HODs, we find excellent recovery of both $A$ and $A_e$, with maximum recovery error significantly less than their $1\sigma$ errorbars quoted in Table~\ref{tab:bestfits}. We also generally get good recovery accuracy ($<1\sigma$ error) on most other HOD parameters, such as $M_\mathrm{cut}$, $M_1$, $\sigma$, $\alpha$, $s$, $s_v$, $\alpha_c$, and $s_p$. However, the recovery accuracy on $\kappa$ is a notably worse, at around 2$\sigma$. This might be attributed to the fact that the redshift-space 2PCF is not particularly sensitive to changes to $\kappa$, as it only modifies satellite occupation at small halo mass. 
Overall, our tests show that at the current level of systematics, the two assembly bias parameters are well constrained by the redshift-space 2PCF on the scales that we chose. Most of the other extended HOD parameters are also well constrained relative to their errorbars, except for $\kappa$. 

\subsection{Fitting the redshift-space multipoles}
\label{subsec:other_studies_rsd}

While in this work we adopted the novel approach of directly fitting the 2D $\xi(r_p, \pi)$, the more common approach is to fit the first multipoles of the redshift-space 2PCF. In principle, the full multipole expansion should contain the same information as the 2D $\xi(r_p, \pi)$. However, different choices in binning and the fact that most fits were done with only the first two or three multipole terms mean that different regions of the $(r_p, \pi)$ separation space enter the multipole fit with different weights. Thus, the first multipoles capture a different set of clustering information compared to $\xi(r_p, \pi)$. 


The most relevant work is \citet{2015aGuo}, where the authors achieved a good fit on a set of BOSS CMASS redshift-space multipoles without invoking any assembly bias prescription in their HOD, seemingly contradicting our finding. 
Besides that fact that our study chooses a different data compression in $\xi(r_p, \pi)$, another key difference between \citet{2015aGuo} and this study lies in our novel velocity bias model. 
While our velocity bias model changes the satellites' velocities and positions simultaneously to preserve Newtonian physics, the \citet{2015aGuo} model modifies the satellite velocities without changing their radial positions, allowing for exotic satellite trajectories that do not obey the physics of the potential well. It is possible that this extra flexibility in satellite velocity removes the need for assembly bias in the HOD model. To test this hypothesis, we perform $\xi(r_p, \pi)$ fits, replacing our velocity bias model with that of \citet{2015aGuo}. However, we continue to find strong evidence for assembly bias, where the inclusion of assembly bias parameter $A$ improves the $\chi^2$/d.o.f by 1.

This suggests that the different data compression ($\xi(r_p, \pi)$ vs multipoles) might be responsible for the contradicting conclusions regarding assembly bias. To test this, we fit the first multipoles, up to $l = 4$, adopting the velocity bias model of \citet{2015aGuo}. Indeed, we find that, in this case, the inclusion of $A$ improves the fit by a much smaller amount ($\Delta\chi^2$/d.o.f = 0.2), yielding weak evidence for assembly bias. Furthermore, we showcase the best-fit multipoles in Figure~\ref{fig:multipole_fits}. We see that the best-fit manages to reproduce the data up to $l = 6$, but fails to reproduce the data at $l = 8$. This suggests that perhaps the first multipoles does not capture the full redshift-space clustering information at these scales, and there is a significant amount of information leftover in the high multipoles. This might not be surprising considering the small-scale RSD signature is dominated by pairs with small transverse separation ($r_p \sim 1$Mpc) and large LOS separation ($\pi\sim 20$Mpc). Such signature is poorly localized in a multipole decomposition. We also examine the multipoles of the $\xi(r_p, \pi)$ fit and find that while it does not reproduce the data at $l = 2, 4$ quite as well as the multipole fit, it does reproduce the data much better at $l = 8$. Thus, this shows that the multipoles capture a different subset of the full clustering information than $\xi(r_p, \pi)$, resulting in different HOD fits. 


\citet{2016Saito} implements a sub-halo abundance matching (SHAM) model based on the peak maximum circular velocity $V_{\mathrm{peak}}$ to model the BOSS CMASS 2PCF. The $V_{\mathrm{peak}}$-based SHAM model is of particular interest here because it naturally accounts for some assembly bias as $V_{\mathrm{peak}}$ is a derivative of the halo merger tree and thus encodes some assembly history information. However, while they found a good fit on the projected 2PCF, they found poor consistency with the first and second redshift-space multipoles. This again highlights how different compressions of the full redshift-space clustering capture different subset of the information content, and also the lack of constraining power of the projected 2PCF. 

\subsection{Alternative environment definitions and scale dependence}
\label{subsec:rmax}

\begin{figure*}
    \centering
    \hspace*{-0.6cm}
    \subfigure[Tophat environment definition]{
    \includegraphics[width = 3.5in]{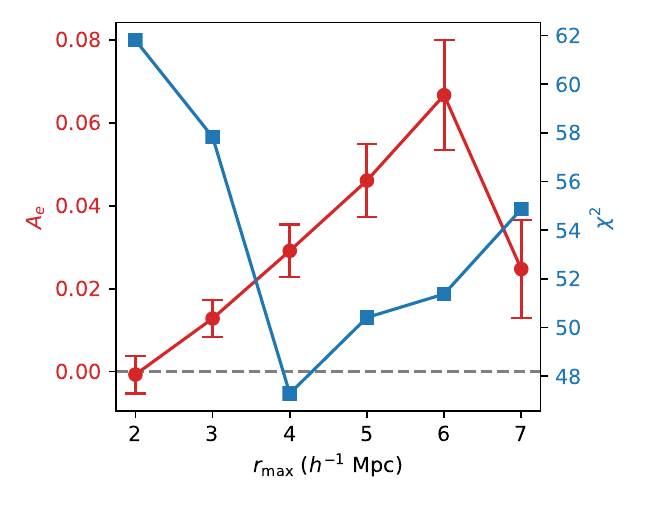}
    }
    \subfigure[Gaussian environment definition]{
    \includegraphics[width = 3.5in]{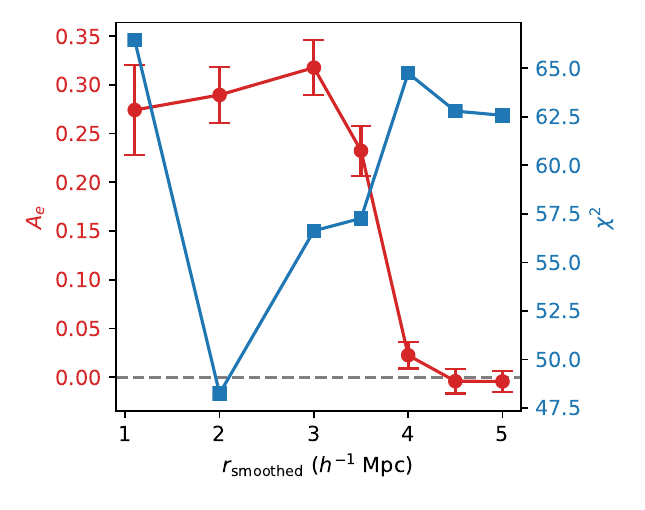}
    }
    \vspace{-0.3cm}
    \caption{The scale dependence of the $A_e$ fit. (a) The best-fit environmental assembly bias parameter $A_e$ and the corresponding best-fit $\chi^2$ as a function of maximum radius of the environment $r_\mathrm{max}$. (b) The same as (a), except the environment is now defined in terms of the local halo density field smoothed with a Gaussian of scale $r_\mathrm{smoothed}$. The error bars represent 1$\sigma$ uncertainties. We see that, in the tophat case, a positive $A_e$ is only preferred by the data when $r_\mathrm{max} \approx 4-6 h^{-1}$Mpc, and in the smoothed case, a positive $A_e$ is only preferred when $r_\mathrm{smoothed} \approx 2 h^{-1}$Mpc. }
    \label{fig:Ae_renv}
\end{figure*}

In this section, we further explore the environmental assembly bias by testing variations to the definition of halo environment. Specifically, we vary the radius, $r_\mathrm{max}$, within which the local environment is calculated. Then, we test a different definition of the local environment altogether, where we use the Gaussian smoothed local density field instead of mass enveloped within $r_\mathrm{max}$. 

The left panel of Figure~\ref{fig:Ae_renv} shows the best-fit environmental assembly bias parameter $A_e$ and the corresponding best-fit $\chi^2 $when we vary the maximum radius $r_\mathrm{max}$ of the halo environment definition. Note again that the environment is defined as the sum of the mass of neighboring halos beyond the virial radius but within $r_\mathrm{max}$. We refer to this mass definition as the tophat definition. We see that the amplitude of the $A_e$ best-fit peaks at $r_\mathrm{max} = 6 h^{-1}$Mpc whereas the smallest $\chi^2$ is achieved at $r_\mathrm{max} = 4 h^{-1}$Mpc. At smaller and larger $r_\mathrm{max}$, the best-fit $A_e$ trends towards zero and the $\chi^2$ increases. Qualitatively, this shows that the goodness-of-fit of the model is sensitive to the choice of $r_\mathrm{max}$, specifically we find $r_\mathrm{max} \approx 4$-$6 h^{-1}$Mpc to be strongly preferred by the data. 

The right panel shows the best-fit $A_e$ and the corresponding $\chi^2$ for the Gaussian smoothed environment definition, where we define the environment as the local halo number density smoothed with a Gaussian kernel of scale $r_\mathrm{smoothed}$. We see a qualitatively similar behavior as the tophat definition, where the goodness-of-fit of the model is strongly dependent on the choice of $r_\mathrm{smoothed}$. In this case, we find $r_\mathrm{smoothed} \approx 2 h^{-1}$Mpc is the smoothing scale that best minimizes $\chi^2$ and maximizes the $A_e$ value. This is qualitatively consistent with the preference for $r_\mathrm{max} \approx 4$-$6 h^{-1}$Mpc in the tophat case, given the extended nature of a Gaussian filter. We should note that the actual value of $A_e$ is not comparable between the two environment definitions, as the amplitude of $A_e$ is degenerate with the amplitude of the environment parameter $f_\mathrm{env}$ (Equation~\ref{equ:Ae}). Because the Gaussian smoothed environment is defined on the halo neighbor count while the tophat environment is defined on mass, the amplitude and distribution of $f_\mathrm{env}$ are actually quite different between the two definitions. 

Both panels suggest that the halo environment at some specific scale of a few megaparsecs is particularly informative of the assembly bias effects, at least for the particular observable we are fitting. While it is not clear what phenomena drive this preference for a specific scale, it does vote against explanations that focus on much smaller scales. 

\begin{figure}
    \centering
    \hspace*{-0.6cm}
    \includegraphics[width = 3.5in]{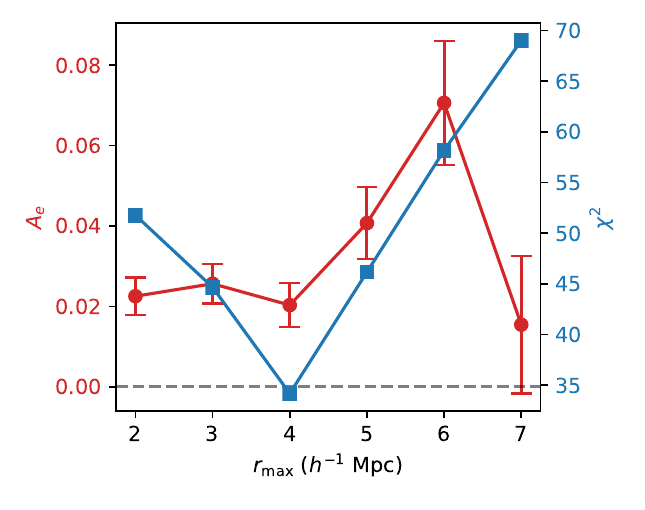}
    \vspace{-0.3cm}
    \caption{The scale dependence of the $A_e$ fit for the enlarged redshift-space 2PCF. The environment is defined with the tophat, same as the left panel of Figure~\ref{fig:Ae_renv}. The enlarged redshift-space 2PCF is computed in 8 log-spaced bins from 0.338$h^{-1}$Mpc to 60$h^{-1}$Mpc in the transverse direction and 6 linearly-spaced bins from 0 to 60$h^{-1}$Mpc in the LOS direction. We see the same behavior as the left panel of Figure~\ref{fig:Ae_renv}, with the amplitude of the $A_e$ best-fit peaking at $r_\mathrm{max} = 6 h^{-1}$Mpc and the $\chi^2$ minimized at $r_\mathrm{max} = 4 h^{-1}$Mpc.}
    \label{fig:Ae_renv_2x}
\end{figure}

It is possible that the preference for specific $r_\mathrm{max}$ is a result of the scales that our redshift-space 2PCF is defined on, specifically from 0.169$h^{-1}$Mpc to 30$h^{-1}$Mpc in the transverse direction and from 0 to 30$h^{-1}$Mpc in the LOS direction. As shown in Figure~4 of \citet{2020Xu}, environment defined at different scales dramatically change its signature on galaxy clustering. Thus, it is possible that the same observable at a different scale would prefer environment defined within a different radius. To test this effect, we generate the redshift-space 2PCF at twice the scale, specifically in 8 log-spaced bins from 0.338$h^{-1}$Mpc to 60$h^{-1}$Mpc in the transverse direction and 6 linearly-spaced bins from 0 to 60$h^{-1}$Mpc in the LOS direction. We fit this enlarged redshift-space 2PCF using our extended HOD with the tophat environment definition and show the best-fit values of $A_e$ and the corresponding $\chi^2$ in Figure~\ref{fig:Ae_renv_2x}.
We see the same behavior as the left panel of Figure~\ref{fig:Ae_renv} despite doubling the scales of the 2PCF, with the amplitude of $A_e$ peaking at $r_\mathrm{max} = 6 h^{-1}$Mpc and the $\chi^2$ minimized at $r_\mathrm{max} = 4 h^{-1}$Mpc. This suggests that the scale preference in $r_\mathrm{max}$ is not a result of the scales imprinted in the 2PCF bins. We believe that this serves as further evidence that the local environment, specifically defined with an $r_\mathrm{max} = 4$-$6 h^{-1}$Mpc, is a physically meaningful indicator of assembly bias and is likely tracing underlying processes happening at a few megaparsecs around halos that truly drive the assembly bias signature. 

It is beyond the scope of this paper to explore the exact underlying processes that drive the environmental assembly bias signature, but we speculate that galaxies and halos in dense environments undergo processes such as mergers, tidal disruptions, and feedback processes, whereas galaxies in under-dense environments evolve mostly passively. These processes can lead to environment dependence in halo occupation. \citet{2020Amodeo} found observational evidence of baryons being expelled beyond the halo virial radius due to a number of feedback effects. The same study also found that this effect could account for up to 50$\%$ of the lensing tension. Another likely relevant phenomenon is the splashback radius \citep{2014Diemer, 2014Adhikari, 2015More}, where studies have found that a traditional halo boundary definition such as $r_\mathrm{virial}$ excludes some gravitationally bound subhalos on highly eccentric orbits. Other parallel studies have also suggested that a more physical halo radius is ${\sim}2$-$3$ times larger than the traditional virial radius \citep{2014Wetzel, 2015Wetzel, 2016Sunayama}. In the context of HOD models, it might be better to consider these splashback halos as part of the host halo instead of as individual halos in the vicinity of larger halos. The fact that we get an extreme best-fit value for $s_p$, which puts some satellites on highly eccentric orbits, might further be indicative of splashback halos. It is possible that the incorporation of splashback radius could alleviate the need for the environmental assembly bias.  In fact, \citet{2020Mansfield} found that the splashback radius can account for much, though not all, of the halo assembly bias signature in simulations. 
 
The importance of halo environment and the recent pushes to enlarge halo boundaries might also point to shortcomings of the halo model overall. In the context of galaxy-halo connection models, instead of populating galaxies per halo, we might get closer to the true galaxy distribution by populating galaxies in larger groups, which are loosely defined as a group of closely associated halos and their local environment. Such a group-based galaxy occupation model, if correctly defined, could eliminate the need to account for local environment and to redefine halo radii. It can significantly simplify existing extended HOD models, such as the one we used in this paper. We defer an exploration of this topic to a future paper.

\subsection{Comparisons to previous studies on environment-based HOD}
\label{subsec:other_studies}

Two previous papers, \citet{2020Hadzhiyska} and \citet{2020Xu}, systematically tested the effectiveness of various secondary HOD dependences in capturing the assembly bias signature, \citet{2020Hadzhiyska} through hydrodynamical simulations and \citet{2020Xu} through semi-analytical models. While they both found the halo environment to be an excellent indicator of assembly bias, there are some key differences between our work and the two previous papers. 

In terms of galaxy samples, in this work we are focusing on LRGs, whereas both previous works focused on much less massive $L_\star$-type galaxies. \citet{2020Hadzhiyska} considered a mass-selected sample of $L_\star$-type galaxies with a number density of $1.3\times 10^{-3} h^3$Mpc$^{-3}$, an order of magnitude higher than our LRG sample. \citet{2020Xu} similarly looked at three samples of mass-selected $L_\star$-type galaxies of density $n_1 = 0.00316 h^3$Mpc$^{-3}$, $n_2 = 0.01 h^3$Mpc$^{-3}$, and $n_3 = 0.0316 h^3$Mpc$^{-3}$, which correspond to stellar-mass thresholds of $3.88\times 10^{10} h^{-1} M_\odot$, $1.42\times 10^{10} h^{-1} M_\odot$, and $0.185 \times 10^{10} h^{-1} M_\odot$, respectively. In comparison, CMASS LRGs are believed to have a typical stellar mass of a few times $10^{11} M_\odot$ \citep{2013Maraston}. 

The definition of halo environment is also different across all three works. Our work largely inherits the environment definition from \citet{2020Hadzhiyska} with minor differences, namely the normalized enclosed mass of subhalos within 5$h^{-1}$Mpc. Most notably, \citet{2020Hadzhiyska} picked the $r_{200m}$ as the inner radius of the environment, while we used the virial radius for convenience. \citet{2020Xu}, however, defined their halo environment as the local mass over-density smoothed with a Gaussian filter, computed with all simulation particles instead of just sub-halos.  They found that the smoothed over-density with a filter scale of $1.25 h^{-1}$Mpc best predicts the assembly bias signature. In a more recent work, \citet{2020bHadzhiyska} adopt an environment definition more similar to that of \citet{2020Xu}, which uses the smoothed matter density field (with a Gaussian kernel of scale $1.1 \ {\rm Mpc}/h$) around a halo to quantify its environment dependence. They show that augmenting the HOD model with a secondary dependence on environment in a manner that matches the two-point clustering, also successfully recovers additional statistical probes of the galaxy distribution such as the g-g lensing signal, the void-galaxy cross-correlation function, and moments of the galaxy density field. In this work, we find a somewhat larger optimal radius, which can be attributed to the different environment definitions and the different galaxy samples. It is also possible that different underlying physical processes drive the environmental assembly bias in L-type galaxies versus in LRGs. Finally, we caution that all these studies are limited by the fidelity of the hydrodynamical simulations and the semi-analytical models used. 



\section{Conclusion}
\label{sec:conclusion}

In this paper, we model the observed redshift-space 2PCF of BOSS CMASS galaxy sample with an extended Hao Occupation Distribution (HOD) model that includes two prescriptions of galaxy assembly bias and other physically motivated additions. We found that while the standard 5-parameter HOD provides a poor fit to the redshift-space 2PCF ($\chi^2 = 151$, d.o.f = 42), the extended HOD achieves a substantially better fit with $\chi^2$ = 50 (d.o.f$=$37) (Table~\ref{tab:bestfits} and Figure~\ref{fig:xi_bestfit}). The redshift-space 2PCF also strongly prefers the simultaneous inclusion of both $A$ and $A_e$, which, respectively, represent the assembly bias associated with halo concentration and halo environment. The preference for HOD models incorporating $A$ and $A_e$ is expressed with their corresponding $\Delta\mathrm{BIC}$: $\Delta \mathrm{BIC} = 19$ for $A$ and $\Delta \mathrm{BIC} = 15$ for $A_e$. The HOD model that includes both assembly biases is strongly favored over an HOD that includes neither, with $\Delta \mathrm{BIC} = 36$. When only one assembly bias term is included, the fit on the redshift-space 2PCF is significantly worse than when both assembly bias terms are included. Our results highlight the importance of a flexible assembly bias and HOD model in accurately modeling redshift-space clustering on the nonlinear scale and expose the deficiencies of the standard 5-parameter HOD model in producing realistic galaxy distributions. 

The best-fit yields a negative concentration-based assembly bias parameter $A$ (${\sim}10\sigma$, preferentially assigning galaxies to puffier less massive halos) and a positive environmental assembly bias parameter $A_e$ (${\sim}3$-$5\sigma$, preferentially assigning galaxies to less massive halos in denser environments). Compared to the projected correlation function, the redshift-space 2PCF pushes both assembly bias parameters in the direction of reducing typical halo mass per galaxy at fixed clustering, showcasing the extra constraining power contained in the LOS structure of the redshift-space 2PCF. Specifically, the inclusion of $A$ decrease the average halo mass per galaxy by $12\%$, whereas the inclusion of $A_e$ decreases the average halo mass by $10\%$. The HOD fit with both assembly biases constrained on redshift-space 2PCF yields $26\%$ lower average halo mass compard to an HOD constrained on the projected correlation function. 

We additionally showed that, by assigning galaxies to lower mass halos, the extended HOD constrained on the redshift-space 2PCF predicts the observed g-g lensing signal to within $1\sigma$ (Figure~\ref{fig:lensing}). This represents a significant improvement compared to predictions by HODs constrained on the projected 2PCF. This result translates the perceived tension between galaxy clustering and lensing to a tension between more informative clustering measurements and over-simplified galaxy-halo connection models. This result again highlights the importance of building more flexible galaxy-halo connection models and utilizing the more informative statistics such as the redshift-space 2PCF. 

We offer strong evidence for including an environmental galaxy assembly bias term in the HOD. In addition to showing how the environmental assembly bias significantly improves the redshift-space 2PCF fit and the lensing prediction, we further showcase the consistency of a positive $A_e$ fit over variations to the fitting routine, variations to the HOD model, and modest perturbations to the cosmology (Figure~\ref{fig:Aefits}). 
Combining this result with previous simulation studies that showed halo environment as an excellent indicator of assembly bias and being able to recover various observables, we believe an environmental assembly bias term is a physical and indispensable addition to the HOD.
We recommend that future studies include environmental assembly bias in their HOD prescriptions in order to construct more realistic galaxy mocks. 

In order to better understand the underlying processes driving the environmental dependence, we tested different maximum radii in the environment definitions, and we found that halo environment defined on scales of around 4-6$h^{-1}$Mpc is preferred by the data (Figure~\ref{fig:Ae_renv}). This scale preference holds even when doubling the bin sizes of the observed 2PCF (Figure~\ref{fig:Ae_renv_2x}). This suggests that the preference for 4-6$h^{-1}$Mpc is not a result of the specific scales chosen when binning the observable, but rather determined by the underlying physical processes that exhibits a similar characteristic scale. We speculate that such processes are intra-halo and might include mergers, tidal-disruptions, and feedback processes. We also suggest that the dependence on environment within 4-6$h^{-1}$Mpc might also point to the need to define halos at a much larger radius, such as the splashback radius.

While we found strong observational evidence for our prescriptions of assembly bias, there is no guarantee that our prescriptions capture the full underlying galaxy assembly bias effect. It is enitrely possible that a better and possibly more simplistic assembly bias prescription can achieve the same or even better fit on the redshift-space galaxy clustering and g-g lensing. 
In upcoming papers, we plan on continuing to explore more physically motivated halo properties as sources of assembly bias, potentially leveraging the halo merger tree and constructing mass-dependent assembly bias prescriptions. We additionally plan on performing a joint analysis on the redshift-space 3PCF, galaxy-galaxy lensing, and the squeezed 3-point correlation function \citep{2017Yuan} with our extended HOD model to derive cosmological constraints.

\section*{Acknowledgements}
We would like to thank Lehman Garrison, Johannes Lange, Jeremy Tinker, Joe DeRose, Andrew Hearin, Martin White, Andrew Zentner, and Josh Speagle for fruitful discussions.
DJE is supported by U.S. Department of Energy grant DE-SC0013718 and as a Simons Foundation Investigator. SB is supported by Harvard University through the ITC Fellowship. HG is supported by the National Science Foundation of China (Nos. 11773049, 11833005, 11828302, 11922305).

\section*{Data Availability Statement}
The simulation data are available at \url{https://lgarrison.github.io/AbacusCosmos/} and \url{https://abacussummit.readthedocs.io/en/latest/}. Researchers wishing to gain access to the extended HOD code can refer to the publicly available \textsc{GRAND-HOD} code at  \url{https://github.com/SandyYuan/GRAND-HOD} or contact the lead author of this paper for details.

\bibliographystyle{aasjournal}
\bibliography{biblio}

\begin{thebibliography}{}
\expandafter\ifx\csname natexlab\endcsname\relax\def\natexlab#1{#1}\fi
\providecommand{\url}[1]{\href{#1}{#1}}

\bibitem[{{Abadi} {et~al.}(2010){Abadi}, {Navarro}, {Fardal}, {Babul}, \&
  {Steinmetz}}]{2010Abadi}
{Abadi}, M.~G., {Navarro}, J.~F., {Fardal}, M., {Babul}, A., \& {Steinmetz}, M.
  2010, \mnras, 407, 435

\bibitem[{{Adhikari} {et~al.}(2014){Adhikari}, {Dalal}, \&
  {Chamberlain}}]{2014Adhikari}
{Adhikari}, S., {Dalal}, N., \& {Chamberlain}, R.~T. 2014, \jcap, 11, 019

\bibitem[{{Alam} {et~al.}(2017){Alam}, {Miyatake}, {More}, {Ho}, \&
  {Mandelbaum}}]{2017bAlam}
{Alam}, S., {Miyatake}, H., {More}, S., {Ho}, S., \& {Mandelbaum}, R. 2017,
  \mnras, 465, 4853

\bibitem[{{Amodeo} {et~al.}(2020){Amodeo}, {Battaglia}, {Schaan}, {Ferraro},
  {Moser}, {Aiola}, {Austermann}, {Beall}, {Bean}, {Becker}, {Bond},
  {Calabrese}, {Calafut}, {Choi}, {Denison}, {Devlin}, {Duff}, {Duivenvoorden},
  {Dunkley}, {D{\"u}nner}, {Gallardo}, {Hall}, {Han}, {Hill}, {Hilton},
  {Hilton}, {Hlo{\v{z}}ek}, {Hubmayr}, {Huffenberger}, {Hughes}, {Koopman},
  {MacInnis}, {McMahon}, {Madhavacheril}, {Moodley}, {Mroczkowski}, {Naess},
  {Nati}, {Newburgh}, {Niemack}, {Page}, {Partridge}, {Schillaci}, {Sehgal},
  {Sif{\'o}n}, {Spergel}, {Staggs}, {Storer}, {Ullom}, {Vale}, {van Engelen},
  {Van Lanen}, {Vavagiakis}, {Wollack}, \& {Xu}}]{2020Amodeo}
{Amodeo}, S., {Battaglia}, N., {Schaan}, E., {et~al.} 2020, arXiv e-prints,
  arXiv:2009.05558

\bibitem[{{Artale} {et~al.}(2018){Artale}, {Zehavi}, {Contreras}, \&
  {Norberg}}]{2018Artale}
{Artale}, M.~C., {Zehavi}, I., {Contreras}, S., \& {Norberg}, P. 2018, \mnras,
  480, 3978

\bibitem[{{Behroozi} {et~al.}(2019){Behroozi}, {Wechsler}, {Hearin}, \&
  {Conroy}}]{2019Behroozi}
{Behroozi}, P., {Wechsler}, R.~H., {Hearin}, A.~P., \& {Conroy}, C. 2019,
  \mnras, 488, 3143

\bibitem[{{Behroozi} {et~al.}(2013){Behroozi}, {Wechsler}, \&
  {Wu}}]{2013Behroozi}
{Behroozi}, P.~S., {Wechsler}, R.~H., \& {Wu}, H.-Y. 2013, \apj, 762, 109

\bibitem[{{Berlind} \& {Weinberg}(2002)}]{2002Berlind}
{Berlind}, A.~A., \& {Weinberg}, D.~H. 2002, \apj, 575, 587

\bibitem[{{Berlind} {et~al.}(2003){Berlind}, {Weinberg}, {Benson}, {Baugh},
  {Cole}, {Dav{\'e}}, {Frenk}, {Jenkins}, {Katz}, \& {Lacey}}]{2003Berlind}
{Berlind}, A.~A., {Weinberg}, D.~H., {Benson}, A.~J., {et~al.} 2003, \apj, 593,
  1

\bibitem[{{Blumenthal} {et~al.}(1984){Blumenthal}, {Faber}, {Primack}, \&
  {Rees}}]{1984Blumenthal}
{Blumenthal}, G.~R., {Faber}, S.~M., {Primack}, J.~R., \& {Rees}, M.~J. 1984,
  \nat, 311, 517

\bibitem[{{Bolton} {et~al.}(2012){Bolton}, {Schlegel}, {Aubourg}, {Bailey},
  {Bhardwaj}, {Brownstein}, {Burles}, {Chen}, {Dawson}, {Eisenstein}, {Gunn},
  {Knapp}, {Loomis}, {Lupton}, {Maraston}, {Muna}, {Myers}, {Olmstead},
  {Padmanabhan}, {P{\^a}ris}, {Percival}, {Petitjean}, {Rockosi}, {Ross},
  {Schneider}, {Shu}, {Strauss}, {Thomas}, {Tremonti}, {Wake}, {Weaver}, \&
  {Wood-Vasey}}]{2012Bolton}
{Bolton}, A.~S., {Schlegel}, D.~J., {Aubourg}, {\'E}., {et~al.} 2012, \aj, 144,
  144

\bibitem[{{Bond} {et~al.}(1991){Bond}, {Cole}, {Efstathiou}, \&
  {Kaiser}}]{1991Bond}
{Bond}, J.~R., {Cole}, S., {Efstathiou}, G., \& {Kaiser}, N. 1991, \apj, 379,
  440

\bibitem[{{Bose} {et~al.}(2019){Bose}, {Eisenstein}, {Hernquist}, {Pillepich},
  {Nelson}, {Marinacci}, {Springel}, \& {Vogelsberger}}]{2019Bose}
{Bose}, S., {Eisenstein}, D.~J., {Hernquist}, L., {et~al.} 2019, \mnras, 490,
  5693

\bibitem[{{Chen} {et~al.}(2017){Chen}, {Ho}, {Mandelbaum}, {Bahcall},
  {Brownstein}, {Freeman}, {Genovese}, {Schneider}, \& {Wasserman}}]{2017Chen}
{Chen}, Y.-C., {Ho}, S., {Mandelbaum}, R., {et~al.} 2017, \mnras, 466, 1880

\bibitem[{{Chua} {et~al.}(2017){Chua}, {Pillepich}, {Rodriguez-Gomez},
  {Vogelsberger}, {Bird}, \& {Hernquist}}]{2017Chua}
{Chua}, K. T.~E., {Pillepich}, A., {Rodriguez-Gomez}, V., {et~al.} 2017,
  \mnras, 472, 4343

\bibitem[{{Contreras} {et~al.}(2019){Contreras}, {Zehavi}, {Padilla}, {Baugh},
  {Jim{\'e}nez}, \& {Lacerna}}]{2019Contreras}
{Contreras}, S., {Zehavi}, I., {Padilla}, N., {et~al.} 2019, \mnras, 484, 1133

\bibitem[{{Croton} {et~al.}(2007){Croton}, {Gao}, \& {White}}]{2007Croton}
{Croton}, D.~J., {Gao}, L., \& {White}, S.~D.~M. 2007, \mnras, 374, 1303

\bibitem[{{Dawson} {et~al.}(2013){Dawson}, {Schlegel}, {Ahn}, {Anderson},
  {Aubourg}, {Bailey}, {Barkhouser}, {Bautista}, {Beifiori}, {Berlind},
  {Bhardwaj}, {Bizyaev}, {Blake}, {Blanton}, {Blomqvist}, {Bolton}, {Borde},
  {Bovy}, {Brandt}, {Brewington}, {Brinkmann}, {Brown}, {Brownstein}, {Bundy},
  {Busca}, {Carithers}, {Carnero}, {Carr}, {Chen}, {Comparat}, {Connolly},
  {Cope}, {Croft}, {Cuesta}, {da Costa}, {Davenport}, {Delubac}, {de Putter},
  {Dhital}, {Ealet}, {Ebelke}, {Eisenstein}, {Escoffier}, {Fan}, {Filiz Ak},
  {Finley}, {Font-Ribera}, {G{\'e}nova-Santos}, {Gunn}, {Guo}, {Haggard},
  {Hall}, {Hamilton}, {Harris}, {Harris}, {Ho}, {Hogg}, {Holder}, {Honscheid},
  {Huehnerhoff}, {Jordan}, {Jordan}, {Kauffmann}, {Kazin}, {Kirkby}, {Klaene},
  {Kneib}, {Le Goff}, {Lee}, {Long}, {Loomis}, {Lundgren}, {Lupton}, {Maia},
  {Makler}, {Malanushenko}, {Malanushenko}, {Mandelbaum}, {Manera}, {Maraston},
  {Margala}, {Masters}, {McBride}, {McDonald}, {McGreer}, {McMahon}, {Mena},
  {Miralda-Escud{\'e}}, {Montero-Dorta}, {Montesano}, {Muna}, {Myers},
  {Naugle}, {Nichol}, {Noterdaeme}, {Nuza}, {Olmstead}, {Oravetz}, {Oravetz},
  {Owen}, {Padmanabhan}, {Palanque-Delabrouille}, {Pan}, {Parejko},
  {P{\^a}ris}, {Percival}, {P{\'e}rez-Fournon}, {P{\'e}rez-R{\`a}fols},
  {Petitjean}, {Pfaffenberger}, {Pforr}, {Pieri}, {Prada}, {Price-Whelan},
  {Raddick}, {Rebolo}, {Rich}, {Richards}, {Rockosi}, {Roe}, {Ross}, {Ross},
  {Rossi}, {Rubi{\~n}o-Martin}, {Samushia}, {S{\'a}nchez}, {Sayres}, {Schmidt},
  {Schneider}, {Sc{\'o}ccola}, {Seo}, {Shelden}, {Sheldon}, {Shen}, {Shu},
  {Slosar}, {Smee}, {Snedden}, {Stauffer}, {Steele}, {Strauss}, {Streblyanska},
  {Suzuki}, {Swanson}, {Tal}, {Tanaka}, {Thomas}, {Tinker}, {Tojeiro},
  {Tremonti}, {Vargas Maga{\~n}a}, {Verde}, {Viel}, {Wake}, {Watson}, {Weaver},
  {Weinberg}, {Weiner}, {West}, {White}, {Wood-Vasey}, {Yeche}, {Zehavi},
  {Zhao}, \& {Zheng}}]{2013Dawson}
{Dawson}, K.~S., {Schlegel}, D.~J., {Ahn}, C.~P., {et~al.} 2013, \aj, 145, 10

\bibitem[{{Diemer} \& {Kravtsov}(2014)}]{2014Diemer}
{Diemer}, B., \& {Kravtsov}, A.~V. 2014, \apj, 789, 1

\bibitem[{{Dragomir} {et~al.}(2018){Dragomir}, {Rodr{\'\i}guez-Puebla},
  {Primack}, \& {Lee}}]{2018Dragomir}
{Dragomir}, R., {Rodr{\'\i}guez-Puebla}, A., {Primack}, J.~R., \& {Lee}, C.~T.
  2018, \mnras, 476, 741

\bibitem[{{Duffy} {et~al.}(2010){Duffy}, {Schaye}, {Kay}, {Dalla Vecchia},
  {Battye}, \& {Booth}}]{2010Duffy}
{Duffy}, A.~R., {Schaye}, J., {Kay}, S.~T., {et~al.} 2010, \mnras, 405, 2161

\bibitem[{{Eisenstein} {et~al.}(2011){Eisenstein}, {Weinberg}, {Agol},
  {Aihara}, {Allende Prieto}, {Anderson}, {Arns}, {Aubourg}, {Bailey},
  {Balbinot}, \& et~al.}]{2011Eisenstein}
{Eisenstein}, D.~J., {Weinberg}, D.~H., {Agol}, E., {et~al.} 2011, \aj, 142, 72

\bibitem[{{Foreman-Mackey} {et~al.}(2013){Foreman-Mackey}, {Hogg}, {Lang}, \&
  {Goodman}}]{2013Foreman}
{Foreman-Mackey}, D., {Hogg}, D.~W., {Lang}, D., \& {Goodman}, J. 2013, \pasp,
  125, 306

\bibitem[{{Gao} {et~al.}(2005){Gao}, {Springel}, \& {White}}]{2005Gao}
{Gao}, L., {Springel}, V., \& {White}, S.~D.~M. 2005, \mnras, 363, L66

\bibitem[{{Gao} \& {White}(2007)}]{2007Gao}
{Gao}, L., \& {White}, S.~D.~M. 2007, \mnras, 377, L5

\bibitem[{{Garrison} {et~al.}(2016){Garrison}, {Eisenstein}, {Ferrer},
  {Metchnik}, \& {Pinto}}]{2016Garrison}
{Garrison}, L.~H., {Eisenstein}, D.~J., {Ferrer}, D., {Metchnik}, M.~V., \&
  {Pinto}, P.~A. 2016, \mnras, 461, 4125

\bibitem[{{Garrison} {et~al.}(2018){Garrison}, {Eisenstein}, {Ferrer},
  {Tinker}, {Pinto}, \& {Weinberg}}]{2018Garrison}
{Garrison}, L.~H., {Eisenstein}, D.~J., {Ferrer}, D., {et~al.} 2018, \apjs,
  236, 43

\bibitem[{{Guo} {et~al.}(2018){Guo}, {Yang}, \& {Lu}}]{2018Guo}
{Guo}, H., {Yang}, X., \& {Lu}, Y. 2018, \apj, 858, 30

\bibitem[{{Guo} {et~al.}(2012){Guo}, {Zehavi}, \& {Zheng}}]{2012Guo}
{Guo}, H., {Zehavi}, I., \& {Zheng}, Z. 2012, \apj, 756, 127

\bibitem[{{Guo} {et~al.}(2015){Guo}, {Zheng}, {Zehavi}, {Dawson}, {Skibba},
  {Tinker}, {Weinberg}, {White}, \& {Schneider}}]{2015aGuo}
{Guo}, H., {Zheng}, Z., {Zehavi}, I., {et~al.} 2015, \mnras, 446, 578

\bibitem[{{Hadzhiyska} {et~al.}(2020{\natexlab{a}}){Hadzhiyska}, {Bose},
  {Eisenstein}, \& {Hernquist}}]{2020bHadzhiyska}
{Hadzhiyska}, B., {Bose}, S., {Eisenstein}, D., \& {Hernquist}, L.
  2020{\natexlab{a}}, arXiv e-prints, arXiv:2008.04913

\bibitem[{{Hadzhiyska} {et~al.}(2020{\natexlab{b}}){Hadzhiyska}, {Bose},
  {Eisenstein}, {Hernquist}, \& {Spergel}}]{2020Hadzhiyska}
{Hadzhiyska}, B., {Bose}, S., {Eisenstein}, D., {Hernquist}, L., \& {Spergel},
  D.~N. 2020{\natexlab{b}}, \mnras, 493, 5506

\bibitem[{Hansen \& Ostermeier(2001)}]{2001Hansen}
Hansen, N., \& Ostermeier, A. 2001, Evolutionary Computation, 9, 159

\bibitem[{{Hartlap} {et~al.}(2007){Hartlap}, {Simon}, \&
  {Schneider}}]{2007Hartlap}
{Hartlap}, J., {Simon}, P., \& {Schneider}, P. 2007, \aap, 464, 399

\bibitem[{{Hearin} {et~al.}(2016){Hearin}, {Zentner}, {van den Bosch},
  {Campbell}, \& {Tollerud}}]{2016Hearin}
{Hearin}, A.~P., {Zentner}, A.~R., {van den Bosch}, F.~C., {Campbell}, D., \&
  {Tollerud}, E. 2016, \mnras, 460, 2552

\bibitem[{{Klypin} {et~al.}(2011){Klypin}, {Trujillo-Gomez}, \&
  {Primack}}]{2011Klypin}
{Klypin}, A.~A., {Trujillo-Gomez}, S., \& {Primack}, J. 2011, \apj, 740, 102

\bibitem[{{Kraljic} {et~al.}(2019){Kraljic}, {Pichon}, {Dubois}, {Codis},
  {Cadiou}, {Devriendt}, {Musso}, {Welker}, {Arnouts}, {Hwang}, {Laigle},
  {Peirani}, {Slyz}, {Treyer}, \& {Vibert}}]{2019Kraljic}
{Kraljic}, K., {Pichon}, C., {Dubois}, Y., {et~al.} 2019, \mnras, 483, 3227

\bibitem[{{Laigle} {et~al.}(2018){Laigle}, {Pichon}, {Arnouts}, {McCracken},
  {Dubois}, {Devriendt}, {Slyz}, {Le Borgne}, {Benoit-L{\'e}vy}, {Hwang},
  {Ilbert}, {Kraljic}, {Malavasi}, {Park}, \& {Vibert}}]{2018Laigle}
{Laigle}, C., {Pichon}, C., {Arnouts}, S., {et~al.} 2018, \mnras, 474, 5437

\bibitem[{Lam {et~al.}(2015)Lam, Pitrou, \& Seibert}]{cite_numba}
Lam, S.~K., Pitrou, A., \& Seibert, S. 2015, in Proceedings of the Second
  Workshop on the LLVM Compiler Infrastructure in HPC, LLVM ’15 (New York,
  NY, USA: Association for Computing Machinery).
\newblock \url{https://doi.org/10.1145/2833157.2833162}

\bibitem[{{Landy} \& {Szalay}(1993)}]{1993Landy}
{Landy}, S.~D., \& {Szalay}, A.~S. 1993, \apj, 412, 64

\bibitem[{{Lange} {et~al.}(2020){Lange}, {Leauthaud}, {Singh}, {Guo}, {Zhou},
  {Smith}, \& {Cyr-Racine}}]{2020Lange}
{Lange}, J.~U., {Leauthaud}, A., {Singh}, S., {et~al.} 2020, arXiv e-prints,
  arXiv:2011.02377

\bibitem[{{Lange} {et~al.}(2019){Lange}, {van den Bosch}, {Zentner}, {Wang},
  {Hearin}, \& {Guo}}]{2019Lange}
{Lange}, J.~U., {van den Bosch}, F.~C., {Zentner}, A.~R., {et~al.} 2019,
  \mnras, 490, 1870

\bibitem[{{Leauthaud} {et~al.}(2016){Leauthaud}, {Bundy}, {Saito}, {Tinker},
  {Maraston}, {Tojeiro}, {Huang}, {Brownstein}, {Schneider}, \&
  {Thomas}}]{2016Leauthaud}
{Leauthaud}, A., {Bundy}, K., {Saito}, S., {et~al.} 2016, \mnras, 457, 4021

\bibitem[{{Leauthaud} {et~al.}(2017){Leauthaud}, {Saito}, {Hilbert},
  {Barreira}, {More}, {White}, {Alam}, {Behroozi}, {Bundy}, {Coupon}, {Erben},
  {Heymans}, {Hildebrandt}, {Mandelbaum}, {Miller}, {Moraes}, {Pereira},
  {Rodr{\'{\i}}guez-Torres}, {Schmidt}, {Shan}, {Viel}, \&
  {Villaescusa-Navarro}}]{2017Leauthaud}
{Leauthaud}, A., {Saito}, S., {Hilbert}, S., {et~al.} 2017, \mnras, 467, 3024

\bibitem[{{Lee} {et~al.}(2018){Lee}, {Primack}, {Behroozi},
  {Rodr{\'\i}guez-Puebla}, {Hellinger}, \& {Dekel}}]{2018Lee}
{Lee}, C.~T., {Primack}, J.~R., {Behroozi}, P., {et~al.} 2018, \mnras, 481,
  4038

\bibitem[{{Levi} {et~al.}(2013){Levi}, {Bebek}, {Beers}, {Blum}, {Cahn},
  {Eisenstein}, {Flaugher}, {Honscheid}, {Kron}, {Lahav}, {McDonald}, {Roe},
  {Schlegel}, \& {representing the DESI collaboration}}]{2013arXiv1308.0847L}
{Levi}, M., {Bebek}, C., {Beers}, T., {et~al.} 2013, arXiv e-prints,
  arXiv:1308.0847

\bibitem[{{Li} {et~al.}(2008){Li}, {Mo}, \& {Gao}}]{2008Li}
{Li}, Y., {Mo}, H.~J., \& {Gao}, L. 2008, \mnras, 389, 1419

\bibitem[{{Mansfield} \& {Kravtsov}(2020)}]{2020Mansfield}
{Mansfield}, P., \& {Kravtsov}, A.~V. 2020, \mnras, 493, 4763

\bibitem[{{Mao} {et~al.}(2018){Mao}, {Zentner}, \& {Wechsler}}]{2018Mao}
{Mao}, Y.-Y., {Zentner}, A.~R., \& {Wechsler}, R.~H. 2018, \mnras, 474, 5143

\bibitem[{{Maraston} {et~al.}(2013){Maraston}, {Pforr}, {Henriques}, {Thomas},
  {Wake}, {Brownstein}, {Capozzi}, {Tinker}, {Bundy}, {Skibba}, {Beifiori},
  {Nichol}, {Edmondson}, {Schneider}, {Chen}, {Masters}, {Steele}, {Bolton},
  {York}, {Weaver}, {Higgs}, {Bizyaev}, {Brewington}, {Malanushenko},
  {Malanushenko}, {Snedden}, {Oravetz}, {Pan}, {Shelden}, \&
  {Simmons}}]{2013Maraston}
{Maraston}, C., {Pforr}, J., {Henriques}, B.~M., {et~al.} 2013, \mnras, 435,
  2764

\bibitem[{{McEwen} \& {Weinberg}(2018)}]{2018McEwen}
{McEwen}, J.~E., \& {Weinberg}, D.~H. 2018, \mnras, 477, 4348

\bibitem[{{More} {et~al.}(2015){More}, {Miyatake}, {Mandelbaum}, {Takada},
  {Spergel}, {Brownstein}, \& {Schneider}}]{2015More}
{More}, S., {Miyatake}, H., {Mandelbaum}, R., {et~al.} 2015, \apj, 806, 2

\bibitem[{{Navarro} {et~al.}(1997){Navarro}, {Frenk}, \& {White}}]{1997Navarro}
{Navarro}, J.~F., {Frenk}, C.~S., \& {White}, S.~D.~M. 1997, \apj, 490, 493

\bibitem[{{Obuljen} {et~al.}(2020){Obuljen}, {Percival}, \&
  {Dalal}}]{2020Obuljen}
{Obuljen}, A., {Percival}, W.~J., \& {Dalal}, N. 2020, arXiv e-prints,
  arXiv:2004.07240

\bibitem[{{Paranjape} {et~al.}(2015){Paranjape}, {Kova{\v{c}}}, {Hartley}, \&
  {Pahwa}}]{2015Paranjape}
{Paranjape}, A., {Kova{\v{c}}}, K., {Hartley}, W.~G., \& {Pahwa}, I. 2015,
  \mnras, 454, 3030

\bibitem[{{Peacock} \& {Smith}(2000)}]{2000Peacock}
{Peacock}, J.~A., \& {Smith}, R.~E. 2000, \mnras, 318, 1144

\bibitem[{{Peirani} {et~al.}(2017){Peirani}, {Dubois}, {Volonteri},
  {Devriendt}, {Bundy}, {Silk}, {Pichon}, {Kaviraj}, {Gavazzi}, \&
  {Habouzit}}]{2017Peirani}
{Peirani}, S., {Dubois}, Y., {Volonteri}, M., {et~al.} 2017, \mnras, 472, 2153

\bibitem[{{Planck Collaboration} {et~al.}(2016){Planck Collaboration}, {Ade},
  {Aghanim}, {Arnaud}, {Ashdown}, {Aumont}, {Baccigalupi}, {Banday},
  {Barreiro}, {Bartlett}, \& et~al.}]{2016Planck}
{Planck Collaboration}, {Ade}, P.~A.~R., {Aghanim}, N., {et~al.} 2016, \aap,
  594, A13

\bibitem[{{Poudel} {et~al.}(2017){Poudel}, {Hein{\"a}m{\"a}ki}, {Tempel},
  {Einasto}, {Lietzen}, \& {Nurmi}}]{2017Poudel}
{Poudel}, A., {Hein{\"a}m{\"a}ki}, P., {Tempel}, E., {et~al.} 2017, \aap, 597,
  A86

\bibitem[{{Pujol} \& {Gazta{\~n}aga}(2014)}]{2014Pujol}
{Pujol}, A., \& {Gazta{\~n}aga}, E. 2014, \mnras, 442, 1930

\bibitem[{{Reid} {et~al.}(2016){Reid}, {Ho}, {Padmanabhan}, {Percival},
  {Tinker}, {Tojeiro}, {White}, {Eisenstein}, {Maraston}, {Ross},
  {S{\'a}nchez}, {Schlegel}, {Sheldon}, {Strauss}, {Thomas}, {Wake}, {Beutler},
  {Bizyaev}, {Bolton}, {Brownstein}, {Chuang}, {Dawson}, {Harding}, {Kitaura},
  {Leauthaud}, {Masters}, {McBride}, {More}, {Olmstead}, {Oravetz}, {Nuza},
  {Pan}, {Parejko}, {Pforr}, {Prada}, {Rodr{\'\i}guez-Torres},
  {Salazar-Albornoz}, {Samushia}, {Schneider}, {Sc{\'o}ccola}, {Simmons}, \&
  {Vargas-Magana}}]{2016Reid}
{Reid}, B., {Ho}, S., {Padmanabhan}, N., {et~al.} 2016, \mnras, 455, 1553

\bibitem[{{Reid} {et~al.}(2014){Reid}, {Seo}, {Leauthaud}, {Tinker}, \&
  {White}}]{2014Reid}
{Reid}, B.~A., {Seo}, H.-J., {Leauthaud}, A., {Tinker}, J.~L., \& {White}, M.
  2014, \mnras, 444, 476

\bibitem[{{Rodr{\'\i}guez-Torres} {et~al.}(2016){Rodr{\'\i}guez-Torres},
  {Chuang}, {Prada}, {Guo}, {Klypin}, {Behroozi}, {Hahn}, {Comparat}, {Yepes},
  {Montero-Dorta}, {Brownstein}, {Maraston}, {McBride}, {Tinker},
  {Gottl{\"o}ber}, {Favole}, {Shu}, {Kitaura}, {Bolton}, {Scoccimarro},
  {Samushia}, {Schlegel}, {Schneider}, \& {Thomas}}]{2016Torres}
{Rodr{\'\i}guez-Torres}, S.~A., {Chuang}, C.-H., {Prada}, F., {et~al.} 2016,
  \mnras, 460, 1173

\bibitem[{{Saito} {et~al.}(2016){Saito}, {Leauthaud}, {Hearin}, {Bundy},
  {Zentner}, {Behroozi}, {Reid}, {Sinha}, {Coupon}, {Tinker}, {White}, \&
  {Schneider}}]{2016Saito}
{Saito}, S., {Leauthaud}, A., {Hearin}, A.~P., {et~al.} 2016, \mnras, 460, 1457

\bibitem[{{Salerno} {et~al.}(2019){Salerno}, {Mart{\'\i}nez}, \&
  {Muriel}}]{2019Salerno}
{Salerno}, J.~M., {Mart{\'\i}nez}, H.~J., \& {Muriel}, H. 2019, \mnras, 484, 2

\bibitem[{{Scoccimarro} {et~al.}(2001){Scoccimarro}, {Sheth}, {Hui}, \&
  {Jain}}]{2001Scoccimarro}
{Scoccimarro}, R., {Sheth}, R.~K., {Hui}, L., \& {Jain}, B. 2001, \apj, 546, 20

\bibitem[{{Sinha} \& {Garrison}(2020)}]{2020Sinha}
{Sinha}, M., \& {Garrison}, L.~H. 2020, \mnras, 491, 3022

\bibitem[{{Skibba} {et~al.}(2011){Skibba}, {van den Bosch}, {Yang}, {More},
  {Mo}, \& {Fontanot}}]{2011Skibba}
{Skibba}, R.~A., {van den Bosch}, F.~C., {Yang}, X., {et~al.} 2011, \mnras,
  410, 417

\bibitem[{{Song} {et~al.}(2020){Song}, {Laigle}, {Hwang}, {Devriendt},
  {Dubois}, {Kraljic}, {Pichon}, {Slyz}, \& {Smith}}]{2020Song}
{Song}, H., {Laigle}, C., {Hwang}, H.~S., {et~al.} 2020, arXiv e-prints,
  arXiv:2009.00013

\bibitem[{{Sunayama} {et~al.}(2016){Sunayama}, {Hearin}, {Padmanabhan}, \&
  {Leauthaud}}]{2016Sunayama}
{Sunayama}, T., {Hearin}, A.~P., {Padmanabhan}, N., \& {Leauthaud}, A. 2016,
  \mnras, 458, 1510

\bibitem[{{Tinker} {et~al.}(2018{\natexlab{a}}){Tinker}, {Hahn}, {Mao}, \&
  {Wetzel}}]{2018bTinker}
{Tinker}, J.~L., {Hahn}, C., {Mao}, Y.-Y., \& {Wetzel}, A.~R.
  2018{\natexlab{a}}, \mnras, 478, 4487

\bibitem[{{Tinker} {et~al.}(2018{\natexlab{b}}){Tinker}, {Hahn}, {Mao},
  {Wetzel}, \& {Conroy}}]{2018Tinker}
{Tinker}, J.~L., {Hahn}, C., {Mao}, Y.-Y., {Wetzel}, A.~R., \& {Conroy}, C.
  2018{\natexlab{b}}, \mnras, 477, 935

\bibitem[{{Tinker} {et~al.}(2017){Tinker}, {Wetzel}, {Conroy}, \&
  {Mao}}]{2017Tinker}
{Tinker}, J.~L., {Wetzel}, A.~R., {Conroy}, C., \& {Mao}, Y.-Y. 2017, \mnras,
  472, 2504

\bibitem[{{van den Bosch} {et~al.}(2005){van den Bosch}, {Weinmann}, {Yang},
  {Mo}, {Li}, \& {Jing}}]{2005vdBosch}
{van den Bosch}, F.~C., {Weinmann}, S.~M., {Yang}, X., {et~al.} 2005, \mnras,
  361, 1203

\bibitem[{{Villarreal} {et~al.}(2017){Villarreal}, {Zentner}, {Mao}, {Purcell},
  {van den Bosch}, {Diemer}, {Lange}, {Wang}, \& {Campbell}}]{2017Villarreal}
{Villarreal}, A.~S., {Zentner}, A.~R., {Mao}, Y.-Y., {et~al.} 2017, \mnras,
  472, 1088

\bibitem[{{Walsh} \& {Tinker}(2019)}]{2019Walsh}
{Walsh}, K., \& {Tinker}, J. 2019, \mnras, 488, 470

\bibitem[{{Wechsler} {et~al.}(2002){Wechsler}, {Bullock}, {Primack},
  {Kravtsov}, \& {Dekel}}]{2002Wechsler}
{Wechsler}, R.~H., {Bullock}, J.~S., {Primack}, J.~R., {Kravtsov}, A.~V., \&
  {Dekel}, A. 2002, \apj, 568, 52

\bibitem[{{Wechsler} \& {Tinker}(2018)}]{2018Wechsler}
{Wechsler}, R.~H., \& {Tinker}, J.~L. 2018, \araa, 56, 435

\bibitem[{{Wechsler} {et~al.}(2006){Wechsler}, {Zentner}, {Bullock},
  {Kravtsov}, \& {Allgood}}]{2006Wechsler}
{Wechsler}, R.~H., {Zentner}, A.~R., {Bullock}, J.~S., {Kravtsov}, A.~V., \&
  {Allgood}, B. 2006, \apj, 652, 71

\bibitem[{{Wetzel} \& {Nagai}(2015)}]{2015Wetzel}
{Wetzel}, A.~R., \& {Nagai}, D. 2015, \apj, 808, 40

\bibitem[{{Wetzel} {et~al.}(2014){Wetzel}, {Tinker}, {Conroy}, \& {van den
  Bosch}}]{2014Wetzel}
{Wetzel}, A.~R., {Tinker}, J.~L., {Conroy}, C., \& {van den Bosch}, F.~C. 2014,
  \mnras, 439, 2687

\bibitem[{{White} \& {Rees}(1978)}]{1978White}
{White}, S.~D.~M., \& {Rees}, M.~J. 1978, \mnras, 183, 341

\bibitem[{{Wibking} {et~al.}(2019){Wibking}, {Salcedo}, {Weinberg}, {Garrison},
  {Ferrer}, {Tinker}, {Eisenstein}, {Metchnik}, \& {Pinto}}]{2019bWibking}
{Wibking}, B.~D., {Salcedo}, A.~N., {Weinberg}, D.~H., {et~al.} 2019, \mnras,
  484, 989

\bibitem[{{Xu} {et~al.}(2020){Xu}, {Zehavi}, \& {Contreras}}]{2020Xu}
{Xu}, X., {Zehavi}, I., \& {Contreras}, S. 2020, arXiv e-prints,
  arXiv:2007.05545

\bibitem[{{Yoshikawa} {et~al.}(2003){Yoshikawa}, {Jing}, \&
  {B{\"o}rner}}]{2003Yoshikawa}
{Yoshikawa}, K., {Jing}, Y.~P., \& {B{\"o}rner}, G. 2003, \apj, 590, 654

\bibitem[{{Yuan} {et~al.}(2017){Yuan}, {Eisenstein}, \& {Garrison}}]{2017Yuan}
{Yuan}, S., {Eisenstein}, D.~J., \& {Garrison}, L.~H. 2017, \mnras, 472, 577

\bibitem[{{Yuan} {et~al.}(2018){Yuan}, {Eisenstein}, \& {Garrison}}]{2018Yuan}
---. 2018, \mnras, 478, 2019

\bibitem[{{Yuan} {et~al.}(2020){Yuan}, {Eisenstein}, \& {Leauthaud}}]{2019Yuan}
{Yuan}, S., {Eisenstein}, D.~J., \& {Leauthaud}, A. 2020, MNRAS, 493, 5551

\bibitem[{{Zehavi} {et~al.}(2018){Zehavi}, {Contreras}, {Padilla}, {Smith},
  {Baugh}, \& {Norberg}}]{2018Zehavi}
{Zehavi}, I., {Contreras}, S., {Padilla}, N., {et~al.} 2018, \apj, 853, 84

\bibitem[{{Zentner}(2007)}]{2007Zentner}
{Zentner}, A.~R. 2007, International Journal of Modern Physics D, 16, 763

\bibitem[{{Zentner} {et~al.}(2005){Zentner}, {Berlind}, {Bullock}, {Kravtsov},
  \& {Wechsler}}]{2005Zentner}
{Zentner}, A.~R., {Berlind}, A.~A., {Bullock}, J.~S., {Kravtsov}, A.~V., \&
  {Wechsler}, R.~H. 2005, \apj, 624, 505

\bibitem[{{Zentner} {et~al.}(2019){Zentner}, {Hearin}, {van den Bosch},
  {Lange}, \& {Villarreal}}]{2019Zentner}
{Zentner}, A.~R., {Hearin}, A., {van den Bosch}, F.~C., {Lange}, J.~U., \&
  {Villarreal}, A. 2019, \mnras, 485, 1196

\bibitem[{{Zentner} {et~al.}(2014){Zentner}, {Hearin}, \& {van den
  Bosch}}]{2014Zentner}
{Zentner}, A.~R., {Hearin}, A.~P., \& {van den Bosch}, F.~C. 2014, \mnras, 443,
  3044

\bibitem[{{Zhao} {et~al.}(2009){Zhao}, {Jing}, {Mo}, \&
  {B{\"o}rner}}]{2009Zhao}
{Zhao}, D.~H., {Jing}, Y.~P., {Mo}, H.~J., \& {B{\"o}rner}, G. 2009, \apj, 707,
  354

\bibitem[{{Zhao} {et~al.}(2003){Zhao}, {Mo}, {Jing}, \&
  {B{\"o}rner}}]{2003Zhao}
{Zhao}, D.~H., {Mo}, H.~J., {Jing}, Y.~P., \& {B{\"o}rner}, G. 2003, \mnras,
  339, 12

\bibitem[{{Zheng} {et~al.}(2007){Zheng}, {Coil}, \& {Zehavi}}]{2007bZheng}
{Zheng}, Z., {Coil}, A.~L., \& {Zehavi}, I. 2007, \apj, 667, 760

\bibitem[{{Zheng} \& {Weinberg}(2007)}]{2007Zheng}
{Zheng}, Z., \& {Weinberg}, D.~H. 2007, \apj, 659, 1

\bibitem[{{Zheng} {et~al.}(2005){Zheng}, {Berlind}, {Weinberg}, {Benson},
  {Baugh}, {Cole}, {Dav{\'e}}, {Frenk}, {Katz}, \& {Lacey}}]{2005Zheng}
{Zheng}, Z., {Berlind}, A.~A., {Weinberg}, D.~H., {et~al.} 2005, \apj, 633, 791

\bibitem[{{Zhu} {et~al.}(2006){Zhu}, {Zheng}, {Lin}, {Jing}, {Kang}, \&
  {Gao}}]{2006Zhu}
{Zhu}, G., {Zheng}, Z., {Lin}, W.~P., {et~al.} 2006, \apjl, 639, L5

\bibitem[{{Zu}(2020)}]{2020Zu}
{Zu}, Y. 2020, arXiv e-prints, arXiv:2010.01143

\end{thebibliography}
\appendix

\section{M\lowercase{ultipole fits }}
\begin{figure*}
    \centering
    \hspace*{-0.6cm}
    \includegraphics[width = 6.5in]{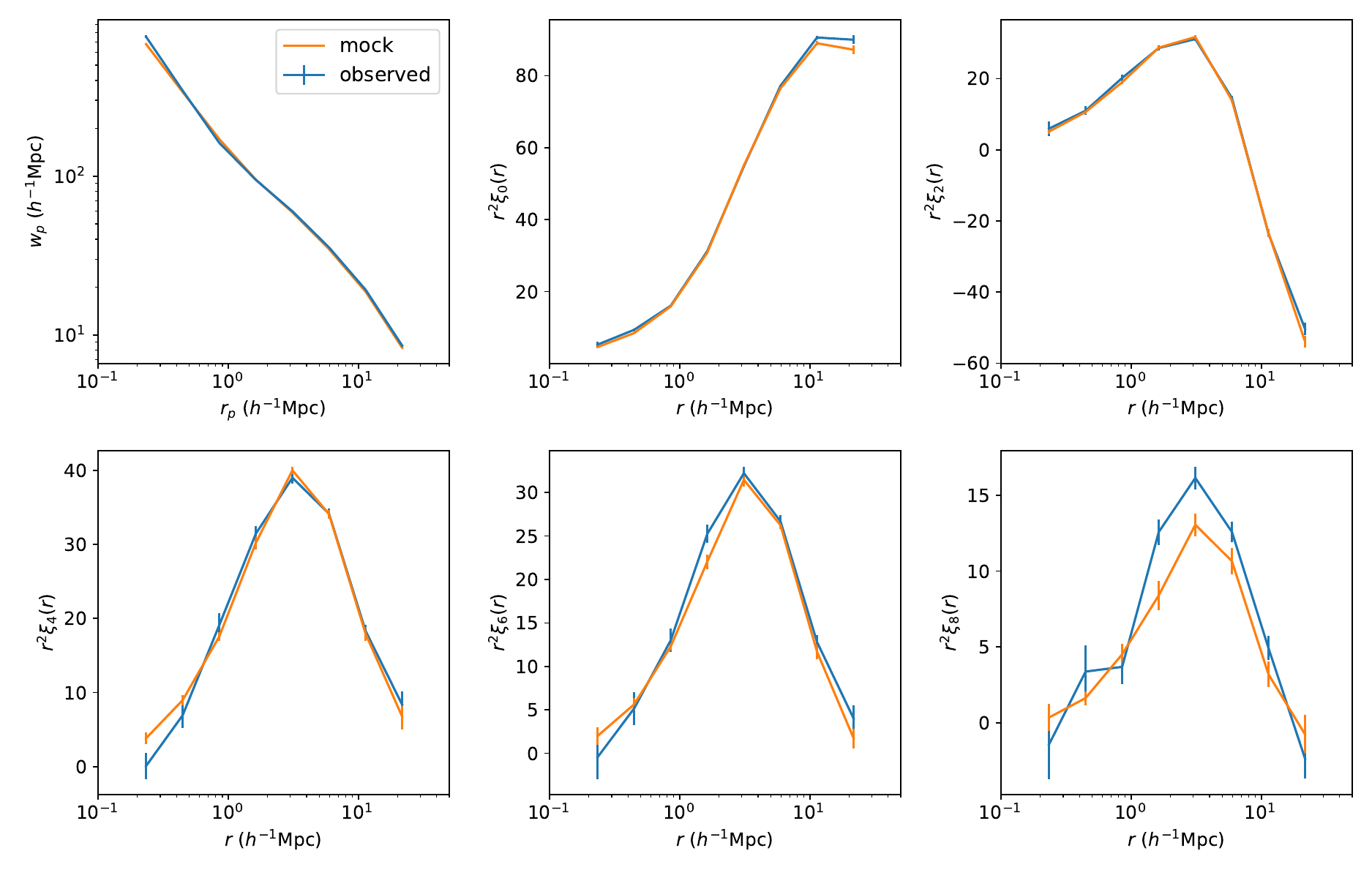}
    \vspace{-0.3cm}
    \caption{The redshift-space multipole fit up to $l = 4$. The fit reproduces the data well up to $l = 6$, but fails to reproduce the data at $l = 8$. This suggests that there is significant information leftover in the higher order multipoles. }
    \label{fig:multipole_fits}
\end{figure*}

In section~\ref{subsec:other_studies_rsd}, we performed a fit on the redshift-space multipoles, adopting the velocity bias model of \citet{2015aGuo}. The resulting best-fit multipoles and projected correlation function $w_p$ are shown in Figure~\ref{fig:multipole_fits}. We see that the multipole fit reproduces the data well up to $l = 6$, but is inconsistent with the data at $l = 8$. This suggests that the first 3 multipoles fail to capture the full information of redshift-space clustering and there is significant amount of information leftover in the higher multipoles.

\label{lastpage}
\end{document}